\documentclass[letterpaper]{article}

\usepackage[T1]{fontenc}

\usepackage{geometry}
\usepackage{setspace}

\usepackage[style = chem-acs,
articletitle = true]{biblatex}
\usepackage{graphicx}
\usepackage{float}
\newfloat{scheme}{htbp}{los}
\floatname{scheme}{Scheme}
\floatname{chart}{Chart}
\newfloat{graph}{htbp}{loh}

\usepackage{chemformula} 
\usepackage[version = 4]{mhchem} 
\usepackage{authblk}
\usepackage{graphicx}
\usepackage{float}
\usepackage{titletoc}
\author[1]{Vaibhav Khanna}
\author[1]{Paul M. Zimmerman*}
\affil[1]{Department of Chemistry, University of Michigan, Ann Arbor, MI 48109, USA}

\title{XC100: A Wavefunction-Derived Exchange-Correlation Energy Dataset for Atomic and Molecular Species}
\date{*Email: paulzim@umich.edu}

\begin{document}

\maketitle

\begin{abstract}
A workflow is introduced for constructing accurate, wavefunction-derived Kohn-Sham (KS) exchange-correlation (XC) energies across atomic and molecular species. Correlated total energies and densities are obtained from configuration interaction wave functions in a polarized triple-zeta basis, cc-pVTZ, and each wavefunction density is then mapped onto a KS determinant. A composite correction adds valence and core basis functions (cc-pVQZ and cc-pCVTZ), plus a two-point Riemann extrapolation to approach the complete-basis-set limit. The workflow is applied to 100 closed-shell atomic and molecular species composed of main group elements to form the XC100 data set. The resulting KS densities closely reproduce the correlated densities, with a median $L_2/N_e$ difference of $1.42\times10^{-5}$, while the composite corrections recover substantial correlation energy beyond the cc-pVTZ reference. Comparison with conventional and machine-learned density functional approximations gives insight into the quality of $E_{\mathrm{xc}}$ across different types of models for XC. In all, this paper's workflow demonstrates a practical route for generating XC reference data for the assessment and development of density functionals.
\end{abstract}





\newcommand{\br}{\textbf{r}}
\newcommand{\rr}{r}
\newcommand{\dr}{\,d\br}
\newcommand{\vxc}{v_{\text{XC}}}
\newcommand{\Exc}{E_{\text{XC}}}
\newcommand{\exc}{e_{\text{XC}}}
\newcommand{\vext}{v_{\text{ext}}}
\newcommand{\veff}{v_{\text{eff}}}
\newcommand{\vH}{v_{\text{H}}}
\newcommand{\vKS}{v_{\text{KS}}}
\newcommand{\rhod}{\rho_{\text{data}}}
\newcommand{\rhoKS}{\rho^{\text{KS}}}
\newcommand{\rhoWF}{\rho^{\text{WF}}}
\newcommand{\bPsi}{\boldsymbol{\Phi}}
\newcommand{\phiKS}{\phi^{\text{KS}}}

\section{Introduction}

Density functional theory (DFT) in the Kohn-Sham (KS) formulation provides a theoretically-motivated and computationally tractable framework for predicting the electronic structure of molecules and materials. \cite{Hohenberg_Kohn,DFT_1,burke2012,Rocky_DFT_Becke} 
DFT's computational advantage is that it is a single-particle theory, where quantum many-body effects are described by the (unknown) exchange-correlation (XC) functional, $E_{xc}[\rho]$. Since the exact form of this functional is unknown, practical calculations require density functional approximations (DFAs). Modern KS DFA development increasingly relies on reference data to parameterize functional forms, train models, and evaluate their transferability across chemical space. \cite{Kaplan2023,Gordon_Exc,nagai2020,kanungo2025}

Reference data used for functional development and assessment have been obtained from experimental measurements and accurate \textit{ab initio} calculations based on wavefunction theory (WFT). These data are assembled into extensive and chemically diverse benchmark collections, covering energies and molecular properties\cite{Interaction_energy_dataset_MHG}. For example, the W4-11 dataset provides precise atomization energies for 140 first- and second-row species,\cite{Karton2011} and GMTKN55 established a broad benchmark of 1505 relative energies spanning main-group thermochemistry, kinetics, and noncovalent interactions.\cite{Goerigk2017} Other examples are MGCDB84,\cite{Gordon_Exc} ACCDB,\cite{Morgante2019} and GSCDB137. The GSCDB137 set contains 8377 reference values across 137 data sets, including transition-metal chemistry, electric-field responses, and vibrational frequencies.\cite{Liang2025} Other efforts have also broadened elemental coverage: TAE-PTComp comprises 2,097 closed-shell, single-reference molecules across much of the periodic table.\cite{Dahl2026} The scale of reference data has likewise continued to increase: MSR-ACC/TAE25 contains 73,040 total atomization energies,\cite{Ehlert2026} and machine-learning XC models can draw on hundreds of thousands of energies.\cite{Luise2026} These resources have been central to the training, parametrization, and validation of density functionals. \cite{Morgante2019,Liang2025,Ehlert2026,Luise2026} Nevertheless, their contents remain dominated by total energies, relative energies, and molecular properties. 

Beyond energetic performance, several studies have examined electron densities produced by DFAs. Medvedev and co-workers reported that historical improvements in energetic performance were not necessarily accompanied by improvements in density quality.\cite{Medvedev2017} These findings prompted debate over how density errors should be quantified and whether conclusions drawn from small atomic test sets generalize to broader chemical applications.\cite{Sim2018} Motivated in part by this debate, new benchmark studies introduced reference dipole moments\cite{Hait2018dipole} and spatial spread of the electron density as probes of density quality. \cite{Hait2021} Whether used in training or testing of DFAs, these quantities are important regularizers that can lead to improvements in the ability to accurately treat the electron density. Together with the examples in the prior paragraph, one might note that XC quantities, for instance the $E_{xc}$ modeled in DFAs, are not represented. Values close to KS theory, such as $E_{xc}$ and $v_{xc}$, require more than just accurate wavefunction calculations.\cite{Khanna2026}

Connecting correlated wavefunctions to the corresponding KS description has historically been difficult. Although the mapping is formally well defined, most practical calculations rely on finite basis sets and approximate wavefunctions, which can introduce substantial numerical sensitivity into the recovery of KS quantities from WFT data.\cite{Khanna2026} These difficulties have limited the routine use of WFT-derived KS information. Recently, advances in inverse DFT and related wavefunction-to-DFT approaches have made it possible to obtain accurate KS orbitals, XC potentials, energy densities, and related quantities from correlated wavefunctions. \cite{Wang_Parr,Zhao1994,Wu2003,Kanungo2019,shi_wasserman_2021,Tribedi2023,Kanungo2023,Ryabinkin2015,Cuevas2015,Ospadov2017,vaibhav_OAvxc} Access to these quantities may create new opportunities for functional development because WFT can provide targets that probe the KS description more directly than total or relative energies alone. Accordingly, machine-learning approaches have increasingly targeted the XC functional and potential.\cite{Schmidt2019machine,Zhou2019toward,nagai2020,Chen2025} For example, neural-network local density approximation (LDA) and generalized gradient approximation (GGA) models were recently trained on the density, XC potential, and XC energy of only five atoms and two molecules. \cite{kanungo2025} The resulting neural-network GGA substantially improved total energies and self-consistent densities relative to the GGA functional PBE\cite{PBE_1,PBE_2}, and achieved accuracy comparable to the higher-rung SCAN\cite{SCAN} meta-GGA across 19 dispersion-insensitive relative energy subsets of GMTKN55.

Given the potential utility of XC data in building data-driven DFAs, we became motivated to extend WFT-derived KS reference data to a broader collection of chemical species. This challenge requires computation of accurate reference wavefunctions for polyatomics and inversion of their densities, while also controlling the cost of large basis sets. Fortunately, composite \textit{ab initio} approaches can reduce the first cost by combining tractable calculations to closely approximate a high-level, large-basis result. The composite strategy underlies protocols such as the Gaussian-$n$ and $Wn$ families. \cite{Curtiss1991G2,Curtiss1998G3,Curtiss2007G4,Karton2006,Hatch2025} Gaussian-$n$ methods begin with a high-level correlated calculation in a moderate basis and add lower-cost corrections obtained from less-correlated methods in larger basis sets. 
Similarly, in the W4 method, the Hartree-Fock and valence-correlation contributions are extrapolated using large basis sets, whereas the substantially more expensive post-CCSD(T) contributions are evaluated in smaller basis sets.\cite{Karton2006}
Composite protocols like $Gn$ and $Wn$ have supplied many of the high-accuracy thermochemical reference values used in the development and assessment of DFAs.\cite{Karton2011,Goerigk2017} The same additive principle can be extended to the KS quantities such as $E_{xc}$, were accurate inversion protocols to be applied. The inversion methods introduced in the previous paragraph provide much of what is needed in this regard, though they have yet to be combined with composite approaches.

In this work, we introduce \textit{XC100}, a data set of reference exchange-correlation energies for 100 closed-shell atomic and molecular species, along with a practical procedure for constructing such references from correlated wavefunctions. Total energies and densities are obtained using accurate configuration interaction wavefunctions. KS states are then generated from these densities,\cite{rask2024} giving $E_{xc}$ as well as its exchange and correlation contributions. Larger-basis correlation effects are incorporated through a composite correction strategy. The \textit{XC100} collection spans a useful range of molecular sizes, electron counts, elemental compositions, and bonding environments. By furnishing direct WFT-derived targets for the XC energy, XC100 complements established benchmark databases and provides reference information for the analysis, assessment, and data-driven development of XC functionals.

\section{Methods and Theory}

\subsection{Wavefunction-to-Kohn-Sham Energy Mapping}

The KS total energy is expressed as
\begin{equation}
    E_{\mathrm{tot}}
    = T_s[\{\phi_i^{\mathrm{KS}}\}]
    + E_{\mathrm{ext}}[\rho]
    + E_H[\rho]
    + E_{xc}[\rho]
    + V_{nn},
    \label{eq:ks_total_energy}
\end{equation}
where $T_s[\{\phi_i^{\mathrm{KS}}\}]$ is the non-interacting kinetic energy evaluated from the KS orbitals $\phi_i^{\mathrm{KS}}$, $E_{\mathrm{ext}}[\rho]$ is the electron-nuclear attraction energy, $E_H[\rho]$ is the classical Hartree energy, and $V_{nn}$ is the nuclear repulsion energy. The remaining term, the exchange-correlation (XC) energy, $E_{xc}[\rho]$, contains the quantum electronic interaction energy together with the difference between the interacting and non-interacting kinetic energies.

Due to the difference in kinetic energies just mentioned, a reference $E_{xc}$ cannot be derived from a correlated wavefunction by a simple subtraction of WFT energy components. This is because the interacting kinetic energy, $T^{\mathrm{WF}}$, is not equal to the non-interacting KS kinetic energy, $T_s$. Even when the two descriptions reproduce the same density,
\begin{equation}
    \rho^{\mathrm{KS}}(\mathbf{r})
    =
    \rho^{\mathrm{WF}}(\mathbf{r}),
    \label{eq:density_matching}
\end{equation}
one generally has $T_s \neq T^{\mathrm{WF}}$. The appropriate $T_s$ must therefore be obtained from the KS determinant associated with the target WFT density.\cite{Zhao1994} Once this determinant is available, the XC energy can be evaluated from the KS energy decomposition,
\begin{equation}
    E_{xc}
    = E_{\mathrm{tot}}^{\mathrm{WF}}
    - T_s
    - E_{\mathrm{ext}}
    - E_H
    - V_{nn}.
    \label{eq:exc_decomposition}
\end{equation}

\subsection{Reference Wavefunction Calculations}


The incremental full configuration interaction (iFCI) method approaches the FCI limit within a fixed basis through a many-body expansion of the correlation energy.\cite{ifci_1,ifci_2} Starting from a perfect-pairing reference \cite{Small2012,Pipek1989,Lawler2010,Janesko2022,VanVoorhis2001,Cullen1996,Cooper2007,Boys1960}, the occupied orbital space is partitioned into localized bonding-antibonding orbital pairs that define the individual bodies of the expansion. At the $n$-body level, groups of $n$ orbital pairs, corresponding to $2n$ electrons, are correlated. The iFCI energy is expressed as
\begin{equation}
    E_{\mathrm{iFCI}}
    =
    E_{\mathrm{ref}}
    + \sum_i \epsilon_i
    + \sum_{i<j} \epsilon_{ij}
    + \sum_{i<j<k} \epsilon_{ijk}
    + \cdots ,
    \label{eq:ifci_expansion}
\end{equation}
where $E_{\mathrm{ref}}$ is the reference energy and $\epsilon_X$ is the incremental correlation contribution associated with a set of bodies $X$. Each increment contains only the correlation not already included through its lower-order subsets. See prior works on iFCI for full details of the energy expansion\cite{ifci_1,ifci_2,ifci_rask,Hatch2025,Hatch2025MBBSA}.

The same incremental construction is applied to the one-particle density matrix,
\begin{equation}
\mathbf{P}_{\mathrm{iFCI}}
=
\mathbf{P}_{\mathrm{ref}}
+
\sum_i \Delta\mathbf{P}_{i}
+
\sum_{i<j}\Delta\mathbf{P}_{ij}
+
\sum_{i<j<k}\Delta\mathbf{P}_{ijk}
+
\cdots ,
\label{eq:ifci_density_matrix}
\end{equation}
where $\Delta\mathbf{P}_{X}$ is the density-matrix increment associated with a set of bodies $X$. As with the energy, contributions from the lower-order subsets are removed from each increment, so particle number is always conserved.

As successively higher-body increments are included, both the energy and density approach their FCI limits within the chosen orbital basis. Because the expansion is constructed from localized orbital pairs, the contribution of each increment generally decreases with increasing body order, allowing the expansion to be truncated at comparatively low order. \cite{ifci_1,ifci_2,Hatch2025,Hatch2025MBBSA}

The resulting correlated wavefunctions provide total energies, electron densities, and interacting kinetic energies for wavefunction-to-KS mapping. The individual iFCI increments are solved using heat-bath configuration interaction (HBCI).\cite{holmes2016,sharma2017,li2018,dang2023} The basis sets, maximum many-body expansion orders, and HBCI convergence parameters are reported in the Computational Details section.

\subsection{Kohn-Sham Determinant Optimization}

Evaluation of $E_{xc}$ from Eq.~\ref{eq:exc_decomposition} requires the non-interacting kinetic energy $T_s$ of the KS determinant. Because a correlated wavefunction calculation provides the density but not the KS determinant, a KS inversion is required to arrive at $E_{xc}$. This is carried out as a finite-basis determinant optimization using the procedure of Rask and co-workers.\cite{rask2024} This procedure, which will now be described, is unconventional in that it is done without reference to the XC potential. This simplifies the optimization of the KS state and helps the overall workflow to be computationally tractable for the 100 systems of the XC100 set.

The KS determinant is obtained by minimizing the difference between the target WFT density and the density of a trial determinant while simultaneously minimizing its non-interacting kinetic energy,
\begin{equation}
\bPsi^{\mathrm{KS}}
=
\operatorname*{arg\,min}_{\bPsi}
\int
\left|
\rho_{\bPsi}(\br)
-
\rhoWF(\br)
\right|^2
\dr
+
\lambda
\left\langle
\bPsi
\middle|
\hat{T}
\middle|
\bPsi
\right\rangle
,
\label{eq:ks_orbital_generation}
\end{equation}
where $\rhoWF(\br)$ is the target WFT density, $\rho_{\bPsi}(\br)$ is the density associated with the trial determinant, and $\lambda$ is a scalar weighting parameter for the kinetic-energy term. 

The density associated with the optimized KS orbitals is
\begin{equation}
\rhoKS(\br)
=
2 \sum_{i=1}^{N_{\mathrm{occ}}}
\left|
\phiKS_i(\br)
\right|^2,
\label{eq:ks_density}
\end{equation}
where $N_{\mathrm{occ}}$ is the number of occupied spatial orbitals. The KS orbitals are expanded in a finite atomic-orbital basis $\{\chi_{\mu}(\br)\}$ as
\begin{equation}
\phiKS_i(\br)
=
\sum_{\mu=1}^{M}
C_{\mu i}
\chi_{\mu}(\br),
\label{eq:ks_orbital_expansion}
\end{equation}
where $C_{\mu i}$ are the molecular-orbital expansion coefficients. These coefficients satisfy the orthonormality condition
\begin{equation}
\mathbf{C}^{T}
\mathbf{S}
\mathbf{C}
=
\mathbf{I},
\label{eq:ks_orthonormality}
\end{equation}
where $\mathbf{S}$ is the atomic-orbital overlap matrix. 

All occupied-virtual orbital rotations are optimized collectively using the analytic gradient of the determinant optimization objective. At each optimization iteration, the  orbital coefficients are used to construct the KS density matrix and evaluate the objective in Eq.~\ref{eq:ks_orbital_generation}. The density contribution is evaluated using the four-center overlap tensor
\begin{equation}
K_{\mu\nu,\lambda\sigma}
=
\int
\chi_\mu(\mathbf r)
\chi_\nu(\mathbf r)
\chi_\lambda(\mathbf r)
\chi_\sigma(\mathbf r)
\,d\mathbf r,
\label{eq:four_center_overlap}
\end{equation}
and differentiation of the objective with respect to the KS density matrix gives
\begin{equation}
A_{\mu\nu}
=
2
\sum_{\lambda\sigma}
K_{\mu\nu,\lambda\sigma}
\left(
P_{\lambda\sigma}^{\mathrm{KS}}
-
P_{\lambda\sigma}^{\mathrm{WF}}
\right)
+
\lambda T_{\mu\nu}
,
\label{eq:density_objective_derivative}
\end{equation}
where $\mathbf{P}^{\mathrm{KS}}$ and $\mathbf{P}^{\mathrm{WF}}$ are the KS and WFT density matrices, respectively, and $T_{\mu\nu}$ is the atomic-orbital kinetic-energy matrix.

After transformation of $\mathbf A$ to the molecular-orbital basis, the gradient with respect to a rotation between orbitals $i$ and $a$ is
\begin{equation}
g_{ia}
=
-\left(
n_i A_{ia}
-
n_a A_{ai}
\right),
\label{eq:orbital_rotation_gradient}
\end{equation}
where $n_i$ is the occupation of orbital $i$. For the closed-shell species considered here, rotations between orbitals having the same occupation leave the density unchanged. Consequently, the relevant gradient components correspond to occupied-virtual rotations. This gradient is used to make orthogonal rotations of the KS orbitals until convergence, as detailed in the Computational Details section.

For $M$ atomic-orbital basis functions, the number of unique atomic-orbital pairs is $M(M+1)/2$. The four-center overlap tensor therefore contains
\begin{equation}
\left[\frac{M(M+1)}{2}\right]^2
=
O(M^4)
\label{eq:four_center_tensor_scaling}
\end{equation}
elements. The four-center overlap tensor is constructed once at the beginning of each determinant optimization and stored for reuse throughout the subsequent collective-gradient iterations. Thus, the $O(M^4)$ tensor contraction is a one-time setup cost rather than a quantity that must be recomputed at every optimization step. The remaining orbital transformations scale as $O(M^3)$ or lower. This setup has a computational advantage over the previous optimizer\cite{rask2024}, which had a cost of $O(M^6)$. Furthermore, the new implementation distributes the four-center overlap tensor and its contractions across GPUs, further reducing the computational time required to generate the KS orbitals.

In a finite orbital basis, the correlated WFT density cannot in general be represented by a single determinant in the same basis, because the KS determinant is restricted to integer orbital occupations and therefore has fewer degrees of freedom with which to reproduce the correlated density. \cite{rask2024} The KS determinant optimization consequently replaces exact density reproduction by minimization of the density difference while simultaneously seeking a low-$T_s$ determinant. Rask and co-workers showed that small values of $\lambda$ define a stable regime in which the density error remains low while the kinetic energy varies only weakly,\cite{rask2024} so the resulting KS state is not unduly sensitive to the choice of $\lambda$. The value of $\lambda$ used for XC100 and tests of its numerical sensitivity are reported in the Computational Details section.

\subsection{Evaluation and Decomposition of the Exchange-Correlation Energy}

XC100 contains closed-shell species with unpolarized densities, such that the KS determinant is constructed from doubly occupied spatial orbitals. Once the KS orbitals associated with each target WFT density were obtained, the individual components of the KS energy decomposition were evaluated. The non-interacting kinetic energy is given by
\begin{equation}
T_s
=
-
\sum_{i=1}^{N_{\mathrm{occ}}}
\int
\phi_i^{\mathrm{KS}*}(\br)
\nabla^2
\phi_i^{\mathrm{KS}}(\br)
\,\dr.
\label{eq:ts}
\end{equation}

The remaining terms in the KS energy decomposition, $E_{\mathrm{ext}}$, $E_H$, and $V_{nn}$, were evaluated in their standard forms from the KS density and nuclear geometry. \cite{DFTBook} The exchange-correlation energy was then obtained from Eq.~\ref{eq:exc_decomposition} using the reference WFT total energy and the non-interacting kinetic energy of the optimized KS determinant.

For a closed-shell KS determinant, the exact-exchange energy was evaluated as
\begin{equation}
E_x
=
-
\sum_{i,j=1}^{N_{\mathrm{occ}}}
\iint
\frac{
\phi_i^{\mathrm{KS}*}(\br)
\phi_j^{\mathrm{KS}}(\br)
\phi_j^{\mathrm{KS}*}(\mathbf{r}')
\phi_i^{\mathrm{KS}}(\mathbf{r}')
}{
|\br-\mathbf{r}'|
}
\,\dr\,d\mathbf{r}'.
\label{eq:ex}
\end{equation}
The correlation energy was then obtained from
\begin{equation}
E_c
=
E_{xc}
-
E_x.
\label{eq:ec}
\end{equation}

The kinetic correlation energy was evaluated as
\begin{equation}
T_c
=
T^{\mathrm{WFT}}
-
T_s.
\label{eq:tc}
\end{equation}

\subsection{Composite Correction Scheme}
\begin{figure}[H]
    \centering
    \includegraphics[width=0.45\textwidth]{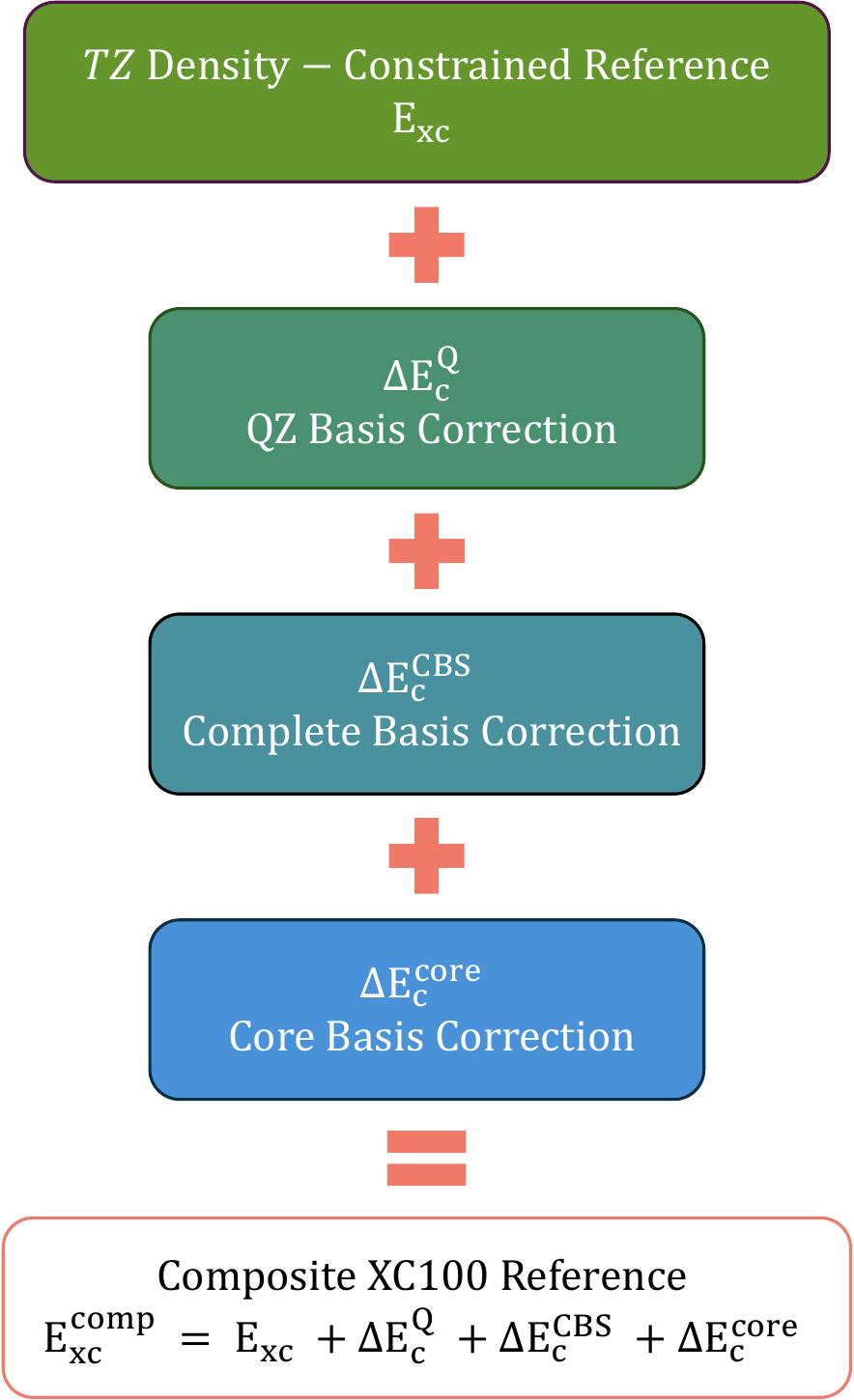}
    \caption{Schematic of the composite correction scheme used to
    construct the XC100 reference energies. The cc-pVTZ
    KS determinant optimization
    provides the baseline reference $E_{xc}^{\mathrm{cc\text{-}pVTZ}}$. Three additive correlation-energy
    corrections are then applied: the cc-pVTZ-to-cc-pVQZ
    finite-basis correction,
    $\Delta E_c^{\mathrm{Q}}$; the remaining
    cc-pVQZ-to-CBS contribution obtained from the Riemann extrapolation,
    $\Delta E_c^{\mathrm{CBS}}$; and the
    cc-pVTZ-to-cc-pCVTZ core-valence basis-set correction,
    $\Delta E_c^{\mathrm{core}}$. Their sum yields the
    composite XC100 reference,
    $E_{xc}^{\mathrm{comp}}$.}
    \label{fig:composite_scheme}
\end{figure}

In a finite basis $B$, a WFT-to-KS mapping yields a finite basis $E_{xc}^{B}$. Obtaining the corresponding quantity in a larger basis would ordinarily require both a larger-basis correlated WFT calculation and the associated KS determinant optimization. A composite alternative is to retain the finite-basis KS reference while estimating the large-basis correction from changes in the WFT correlation energy.
For a WFT-to-KS mapping performed in the cc-pVTZ basis \cite{dunning1989a,prascher2011a,woon1994a}, such a composite XC energy can therefore be defined as
\begin{equation}
E_{xc}^{\mathrm{comp}}
=
E_{xc}^{\mathrm{cc\text{-}pVTZ}}
+
\Delta E_{c}^{\mathrm{cc\text{-}pVQZ}}
+
\Delta E_{c}^{\mathrm{Riemann}}
+
\Delta E_{c}^{\mathrm{cc\text{-}pCVTZ}},
\label{eq:composite_exc}
\end{equation}
where $E_{xc}^{\mathrm{cc\text{-}pVTZ}}$ is obtained from the cc-pVTZ WFT energy and the corresponding KS determinant. The first correction accounts for the change in correlation energy upon increasing the basis from cc-pVTZ to cc-pVQZ,\cite{dunning1989a,prascher2011a,woon1994a} the second estimates the remaining correlation-energy contribution between cc-pVQZ and the complete-basis-set (CBS) limit using a two-point Riemann zeta-function extrapolation\cite{Riemann}, and the third accounts for the change obtained upon replacing the valence-optimized cc-pVTZ basis with the core-valence-optimized cc-pCVTZ basis. \cite{dunning1989a,feller1996a,schuchardt2007a,woon1995a} 
The additive construction of Eq.~\ref{eq:composite_exc} is summarized
schematically in Figure~\ref{fig:composite_scheme}.

For the cc-pVTZ and cc-pVQZ basis-set pair, the two-point Riemann extrapolation factor is
\begin{equation}
A_{4}
=
4^{4}
\left[
\zeta(4)
-
\sum_{\ell=1}^{4}\ell^{-4}
\right],
\label{eq:riemann_factor}
\end{equation}
where $\zeta(s)$ denotes the Riemann zeta function, evaluated here at $s=4$.\cite{Riemann} For a finite-basis correlation energy $E_c^{B}$, the corresponding CBS estimate is
\begin{equation}
E_{c}^{\mathrm{CBS}(T,Q)}
=
E_{c}^{\mathrm{cc\text{-}pVQZ}}
+
A_{4}
\left(
E_{c}^{\mathrm{cc\text{-}pVQZ}}
-
E_{c}^{\mathrm{cc\text{-}pVTZ}}
\right).
\label{eq:riemann_cbs_general}
\end{equation}
Because the cc-pVTZ-to-cc-pVQZ increment is already included separately in Eq.~\ref{eq:composite_exc}, the additional Riemann correction is only the remaining cc-pVQZ-to-CBS contribution,
\begin{equation}
\Delta E_{c}^{\mathrm{Riemann}}
=
E_{c}^{\mathrm{CBS}(T,Q)}
-
E_{c}^{\mathrm{cc\text{-}pVQZ}}.
\label{eq:riemann_correction_general}
\end{equation}

For iFCI wavefunctions, all finite-basis and CBS corrections are constructed from HF-referenced iFCI correlation energies through the two-body level. For a basis $B$, this correlation energy is defined as
\begin{equation}
E_{c,\mathrm{iFCI}}^{(2),B}
=
E_{\mathrm{iFCI}}^{(2),B}
-
E_{\mathrm{HF}}^{B},
\label{eq:ifci_hf_referenced_correlation}
\end{equation}
where the iFCI and HF energies are evaluated in the same basis. The two-body iFCI energy is
\begin{equation}
E_{\mathrm{iFCI}}^{(2),B}
=
E_{\mathrm{ref}}^{B}
+
E_{1\mathrm{c}}^{B}
+
E_{2\mathrm{c}}^{B},
\label{eq:ifci_two_body_total_energy}
\end{equation}
where $E_{\mathrm{ref}}^{B}$ is the perfect-pairing reference energy. Thus, correlation already present in the perfect-pairing reference is included consistently in the finite-basis corrections as well as in the CBS extrapolation.

The cc-pVTZ-to-cc-pVQZ correction is therefore
\begin{equation}
\Delta E_{c}^{\mathrm{cc\text{-}pVQZ}}
=
E_{c,\mathrm{iFCI}}^{(2),\mathrm{cc\text{-}pVQZ}}
-
E_{c,\mathrm{iFCI}}^{(2),\mathrm{cc\text{-}pVTZ}},
\label{eq:ifci_qz_correction}
\end{equation}
and the core-valence basis-set correction is
\begin{equation}
\Delta E_{c}^{\mathrm{cc\text{-}pCVTZ}}
=
E_{c,\mathrm{iFCI}}^{(2),\mathrm{cc\text{-}pCVTZ}}
-
E_{c,\mathrm{iFCI}}^{(2),\mathrm{cc\text{-}pVTZ}}.
\label{eq:ifci_cv_correction}
\end{equation}


The larger-basis corrections in Eqs.~\ref{eq:ifci_qz_correction} and \ref{eq:ifci_cv_correction} therefore correct the two-body HF-referenced correlation energy while retaining the three-body contribution contained in the underlying cc-pVTZ reference $E_{xc}$. No separate basis-set correction is applied to the three-body contribution. This approximation is examined explicitly for \ce{BH3}, \ce{CH4}, and \ce{NH3} by evaluating the three-body correlation contribution in both the cc-pVTZ and cc-pVQZ basis sets (Supporting Information, Table~S2). The absolute cc-pVTZ-to-cc-pVQZ changes are only 0.020-0.148 mHa in total, corresponding to at most approximately 0.015 mHa per electron. The weak basis dependence observed for these representative species supports retaining the three-body contribution from the cc-pVTZ baseline without introducing a separate three-body larger-basis correction.

The same HF-referenced correlation energies are used for the Riemann extrapolation. The extrapolated iFCI correlation energy is
\begin{equation}
E_{c,\mathrm{iFCI}}^{\mathrm{CBS}(T,Q)}
=
E_{c,\mathrm{iFCI}}^{(2),\mathrm{cc\text{-}pVQZ}}
+
A_{4}
\left[
E_{c,\mathrm{iFCI}}^{(2),\mathrm{cc\text{-}pVQZ}}
-
E_{c,\mathrm{iFCI}}^{(2),\mathrm{cc\text{-}pVTZ}}
\right],
\label{eq:ifci_riemann_cbs}
\end{equation}
and the additional Riemann correction entering
Eq.~\ref{eq:composite_exc} is
\begin{equation}
\Delta E_{c}^{\mathrm{Riemann}}
=
E_{c,\mathrm{iFCI}}^{\mathrm{CBS}(T,Q)}
-
E_{c,\mathrm{iFCI}}^{(2),\mathrm{cc\text{-}pVQZ}}.
\label{eq:ifci_riemann_correction}
\end{equation}


For HBCI wavefunctions, the same HF-referenced correlation-energy convention is used throughout, with $E_c^{B}=E_{\mathrm{HBCI}}^{B}-E_{\mathrm{HF}}^{B}$. The corresponding finite-basis and Riemann extrapolation expressions are given in the Supporting Information section titled HBCI Composite-Correction Expressions.

Thus, the cc-pVTZ KS determinant optimization provides the baseline XC energy, while the correlation-energy differences provide the additive larger-basis corrections. The cc-pVQZ calculation supplies the cc-pVTZ-to-cc-pVQZ correlation-energy increment, the Riemann extrapolation supplies the remaining cc-pVQZ-to-CBS correlation-energy tail, and the cc-pCVTZ calculation supplies the core-valence basis-set correction. 

The corrections above are constructed from WFT correlation energies. To express the final reference quantities in the KS-DFT energy decomposition, we now return to the DFT definition $E_c=E_{xc}-E_x$. The composite correlation energy is therefore obtained by subtracting the exact-exchange energy evaluated from the cc-pVTZ optimized KS determinant from the composite XC energy,
\begin{equation}
E_{c}^{\mathrm{comp}}
=
E_{xc}^{\mathrm{comp}}
-
E_{x}^{\mathrm{cc\text{-}pVTZ}}.
\label{eq:composite_correlation_energy}
\end{equation}
Although $E_x$ is also formally evaluated at the cc-pVTZ level, its basis dependence is considerably smaller: full cc-pCVQZ \cite{dunning1989a,woon1995a} inversions for \ce{BH3}, \ce{CH4}, and \ce{NH3} change $E_x$ by at most approximately $0.65$ mHa per electron. Because the additive basis-set corrections modify only the correlation contribution, this expression is equivalently
\begin{equation}
E_{c}^{\mathrm{comp}}
=
E_{c}^{\mathrm{cc\text{-}pVTZ}}
+
\Delta E_{c}^{\mathrm{cc\text{-}pVQZ}}
+
\Delta E_{c}^{\mathrm{Riemann}}
+
\Delta E_{c}^{\mathrm{cc\text{-}pCVTZ}},
\label{eq:composite_correlation_expanded}
\end{equation}
where
\begin{equation}
E_{c}^{\mathrm{cc\text{-}pVTZ}}
=
E_{xc}^{\mathrm{cc\text{-}pVTZ}}
-
E_{x}^{\mathrm{cc\text{-}pVTZ}}.
\label{eq:tz_correlation_energy}
\end{equation}


\section{Computational Details}

For the construction of XC100, geometries for species present in the W4-11 benchmark set were taken directly from the reported structures.\cite{Karton2011} For the remaining species, initial Cartesian coordinates were obtained from the NCI/CADD Chemical Identifier Resolver\cite{NCICADDResolver} and optimized at the restricted B3LYP/cc-pVTZ level using PySCF and the geomeTRIC geometry optimizer.\cite{PySCF,Wang2016Geometric,B3LYP,dunning1989a,prascher2011a,woon1994a} For five Li- and Be-containing species requiring manually constructed starting geometries, a B3LYP/def2-SVP preoptimization was performed before the final B3LYP/cc-pVTZ optimization.\cite{weigend2005a} The final Cartesian coordinates for all 100 species, together with the XC100 energy data and the mapping between system identifiers and chemical names, are available in the XC100 GitHub repository (\url{https://github.com/ZimmermanGroup/XC100}).

Reference wavefunction calculations were performed using HBCI \cite{holmes2016,sharma2017,li2018,dang2023,chien2018} and iFCI.\cite{ifci_1,ifci_2,ifci_rask,Hatch2025,Hatch2025MBBSA} HBCI was used directly for atoms and diatomics because their smaller configuration spaces permit the FCI limit to be approached without introducing the incremental many-body decomposition. HBCI constructs the variational space by selecting determinants that couple most strongly to the current wavefunction and supplements the variational energy with a perturbative correction. Tightening the selection and perturbative thresholds provides a systematic route toward the FCI result. For larger polyatomic species, however, direct HBCI becomes prohibitively expensive. The polynomial-scaling iFCI method was therefore employed to obtain comparable near-FCI accuracy at a reduced computational cost, with HBCI used as the solver for the active-space CI problem associated with each iFCI increment. For atoms and diatomics, direct HBCI calculations employed a variational selection threshold of $\varepsilon_1=5\times10^{-5}$ Ha and a perturbative threshold of $\varepsilon_2=1\times10^{-7}$ Ha. All electrons, including the core electrons, were correlated. For the remaining molecular species, iFCI calculations were performed in a perfect-pairing orbital basis through the three-body, $n=3$, level of the many-body expansion. Previous benchmarks for closed-shell main-group species have shown approximately $1$ mHa precision at this truncation relative to higher-order iFCI results.\cite{ifci_1,ifci_2,Hatch2025MBBSA,Hatch2025} HBCI was used as the solver for each increment with $\varepsilon_1=1\times10^{-4}$ Ha and $\varepsilon_2=1\times10^{-7}$ Ha, and core-electron correlation was included through the two-body level. For each iFCI increment, the virtual orbital space was constructed using the incremental natural orbital (iNO) procedure of Hatch et al.\cite{Hatch2025} Virtual NOs with occupation numbers below $10^{-\zeta}$ were excluded from the correlated orbital space. Three-body increments used this natural-orbital screening with $\zeta=5.5$. Repeating representative three-body calculations with the tighter value $\zeta=8$ changed the iFCI energies by at most $0.02$ mHa (Supporting Information, Table~S5), supporting the use of $\zeta=5.5$. Details of the perfect-pairing orbital construction, natural-orbital generation, and HBCI implementation are provided in the additional computational details section of the Supporting Information.

All reference densities used in the KS determinant optimization were evaluated in the cc-pVTZ basis with the corresponding RI fitting auxiliary basis. \cite{dunning1989a,prascher2011a,woon1994a,hattig2005a,weigend2002a} The basis sets and associated auxiliary bases were obtained from Basis Set Exchange.\cite{pritchard2019a,feller1996a,schuchardt2007a} The non-interacting kinetic-energy weighting parameter was set to $\lambda=1\times10^{-5}$ for all species. Varying $\lambda$ over an order of magnitude, from $1\times10^{-5}$ to $1\times10^{-4}$, changed $E_{\mathrm{xc}}$ by only approximately $0.34$-$0.39$ mHa per electron for \ce{BH3}, \ce{CH4}, and \ce{NH3} (Supporting Information, Table~S3). The KS determinant optimization was performed using an L-BFGS quasi-Newton algorithm and was considered converged when the Euclidean norm of the orbital-rotation gradient was below $10^{-7}$. The reference density, optimized KS determinant, non-interacting kinetic energy $T_s$, exact-exchange energy $E_x$, and kinetic correlation energy $T_c$ were therefore evaluated at the cc-pVTZ level. More details of the determinant optimization are given in the Supporting Information.

Additional reference wavefunction calculations were performed using the cc-pVQZ and cc-pCVTZ basis sets, together with their corresponding RI fitting auxiliary bases,\cite{hattig2005a,weigend2002a,kritikou2015a} to evaluate the additive correlation-energy corrections entering $E_{xc}^{\mathrm{comp}}$. \cite{dunning1989a,prascher2011a,woon1994a,feller1996a,schuchardt2007a,woon1995a} For species treated using iFCI, both larger-basis calculations were truncated after the two-body level. The cc-pVQZ calculations used $\varepsilon_1=2\times10^{-4}$ Ha and $\varepsilon_2=1\times10^{-7}$ Ha, whereas the cc-pCVTZ calculations used $\varepsilon_1=1\times10^{-4}$ Ha and $\varepsilon_2=1\times10^{-7}$ Ha. The slightly looser $\varepsilon_1$ threshold for cc-pVQZ reduces the cost of the larger increment calculations. Varying $\varepsilon_1$ from $1\times10^{-4}$ to $5\times10^{-4}$ Ha changes the representative cc-pVQZ two-body iFCI energies by at most $0.109$ mHa (Supporting Information, Table~S4). Core electrons were correlated through the two-body level in both basis sets. For atoms and diatomics, the cc-pVQZ and cc-pCVTZ HBCI calculations used the same thresholds as the cc-pVTZ calculations, $\varepsilon_1=5\times10^{-5}$ Ha and $\varepsilon_2=1\times10^{-7}$ Ha, with all electrons correlated in each basis.


Self-consistent restricted KS and generalized KS calculations were performed using PySCF\cite{PySCF} in the cc-pVQZ basis. The conventional PW91, PBE, BLYP, SCAN, and B3LYP XC functionals were considered \cite{PW91_1,PW91_2,PBE_1,PBE_2,B88,LYP_1,LYP_2,SCAN,B3LYP}, together with the machine-learned NNGGA functional of Kanungo et al.\cite{kanungo2025} and Skala \cite{Luise2026}. The Skala calculations used the Skala-1.1 model; ``Skala'' refers to this version throughout the remainder of this work. The semilocal functional expressions were evaluated using Libxc.\cite{LibXC} A PySCF level-9 numerical grid with 200 radial and 1454 angular points was used with grid pruning disabled. The self-consistent-field energy convergence threshold was $1\times10^{-10}$ Ha, with a maximum of 500 iterations. Details of the machine-learned functional implementations and their SCF convergence across XC100 are reported in the Supporting Information. All species were treated as closed-shell and spin-unpolarized; the nitrosonium and Zundel cations were assigned charges of $+1$, while all other species were neutral. For the semilocal functionals, $E_{xc}$, $E_x$, and $E_c$ were evaluated separately over the converged self-consistent density. B3LYP was constructed explicitly from its exact-exchange and semilocal components; the precise energy expression and implementation are given in the Supporting Information.

\section{Results and Discussion}

The results are organized around the construction, quality, and characteristics of the XC100 reference data. The wavefunction-to-KS workflow used to generate the reference quantities is introduced first, followed by a description of the chemical scope of the XC100 data set. The accuracy of the KS mapping is then assessed by comparing WF and KS densities. Next, we examine the composite corrections and validate the resulting XC reference energies. Finally, the composite $E_{\mathrm{xc}}$ references are compared to conventional and machine-learned density-functional approximations.

\begin{figure}[H]
    \centering
    \includegraphics[width=0.89\textwidth]{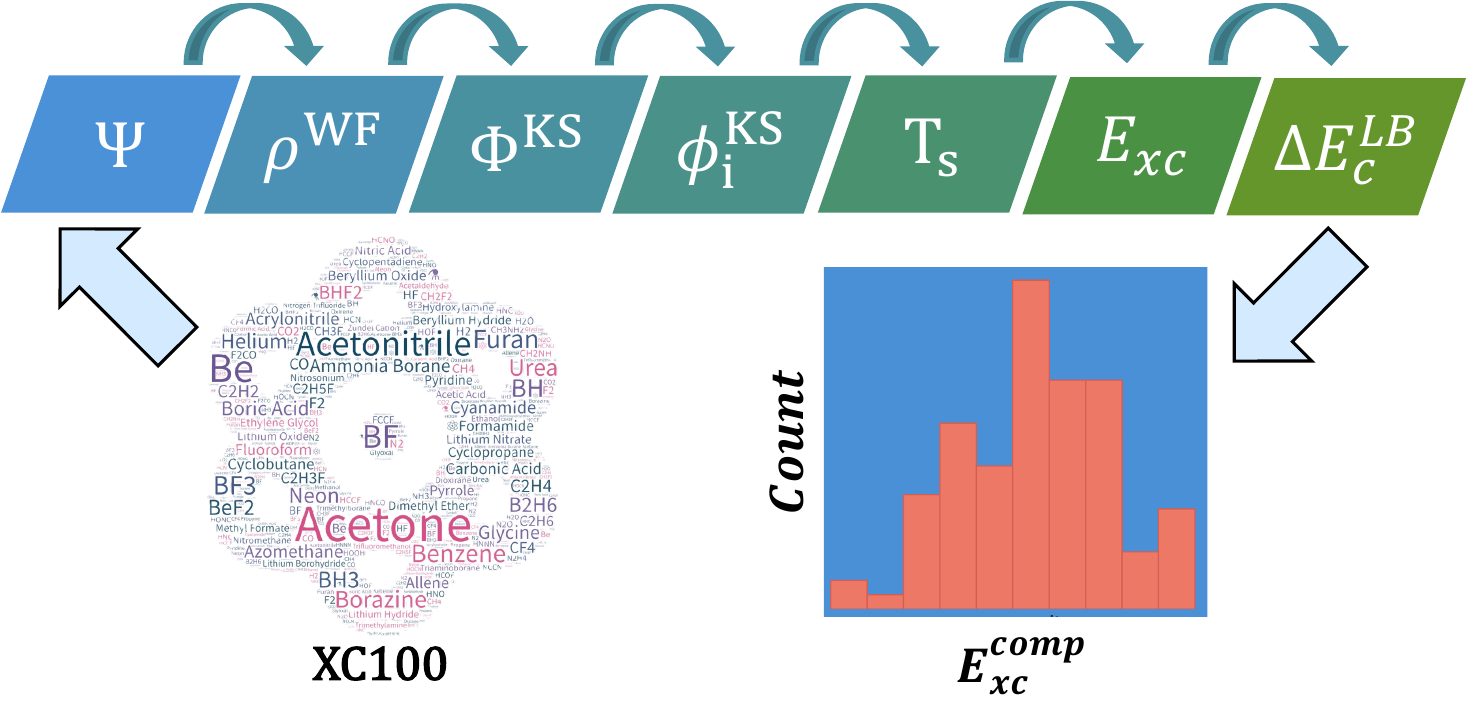}
    \caption{Schematic workflow used to construct XC100. The chemically diverse set of 100 atomic and molecular species is represented by the word cloud. Correlated CI wavefunctions, $\Psi$, provide the reference total energies and densities, $\rho^{\mathrm{WF}}$, which are mapped onto KS determinants, $\Phi^{\mathrm{KS}}$, through equation \ref{eq:ks_orbital_generation}. The resulting KS orbitals $\{\phi_i^{\mathrm{KS}}\}$, provide the non-interacting kinetic energy, $T_s$, required to determine the cc-pVTZ exchange-correlation energy, $E_{xc}$. Additive larger-basis correlation-energy corrections, $\Delta E_c^{\mathrm{LB}}$, comprising the cc-pVQZ, Riemann CBS, and cc-pCVTZ contributions, yield the final composite reference energies, $E_{xc}^{\mathrm{comp}}$.}
    \label{fig:xc100_workflow}
\end{figure}

The workflow used to construct XC100 is summarized in Figure~\ref{fig:xc100_workflow}. For each of the 100 chemical species, a correlated CI wavefunction provides the reference total energy and electron density. The correlated density is then mapped to a KS determinant, providing the KS orbitals required to evaluate $T_s$ and thereby connect the wavefunction energy to the KS energy decomposition. The resulting cc-pVTZ $E_{\mathrm{xc}}$ serves as the baseline reference to which correlation-energy corrections from a larger basis set are subsequently added.
This workflow separates the KS state optimization from the larger-basis treatment of correlation. The accuracy of using the cc-pVTZ KS state as the baseline for this composite construction is explicitly assessed later in this section.

\subsection{Chemical Scope and Composition of XC100}

\begin{figure}[H]
    \centering
    \includegraphics[width=\columnwidth]{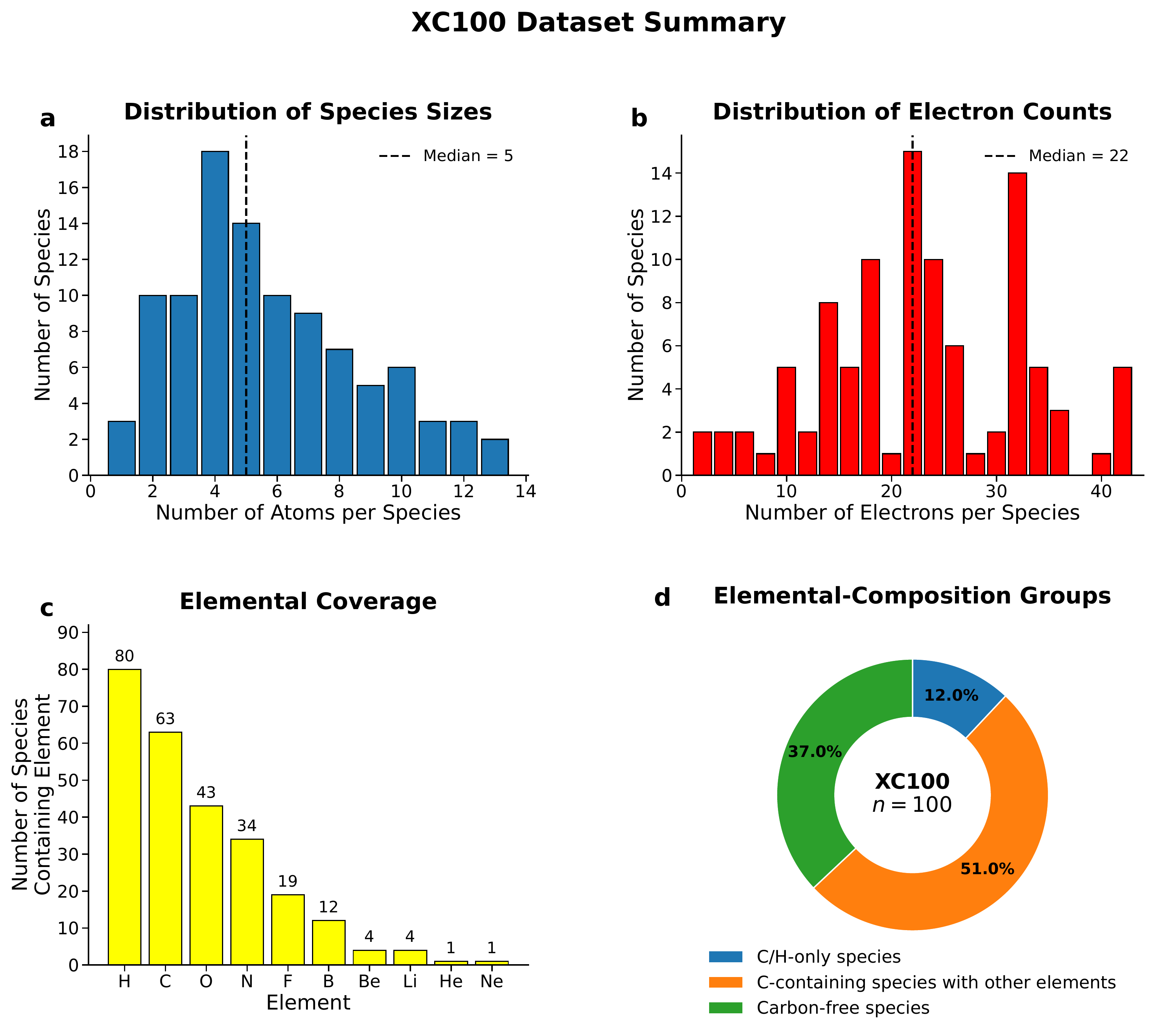}
    \caption{Summary of the chemical composition and size of the XC100 data set. 
    (a) Distribution of the number of atoms per species. 
    (b) Distribution of the number of electrons per species. 
    (c) Number of species containing each chemical element. 
    (d) Classification of the species into C/H-only, carbon plus other heavy element, and carbon-free.}
    \label{fig:xc100_dataset_summary}
\end{figure}

XC100 includes closed-shell atomic and molecular species, as summarized in Figure~\ref{fig:xc100_dataset_summary}. It represents a first step toward extending WFT-derived XC reference data across more diverse classes of electronic structure, including open-shell and strongly-correlated species. The 100 species span a wide range of atom and electron counts, from atoms and diatomics to polyatomic species containing as many as 13 atoms and 42 electrons. The median species contains 5 atoms [Figure~\ref{fig:xc100_dataset_summary}(a)] and 22 electrons [Figure~\ref{fig:xc100_dataset_summary}(b)].

Elemental coverage extends across the first two periods from H through Ne [Figure~\ref{fig:xc100_dataset_summary}(c)]. Hydrogen and carbon are the most frequently represented elements, occurring in 80 and 63 species, respectively, followed by O (43), N (34), F (19), and B (12). Li and Be each occur in four species, while He and Ne provide atomic noble-gas references. This composition samples a range of main-group environments involving heteroatom substitution, ionic and covalent bonding, and multicenter motifs. As one complementary view of this diversity, Figure~\ref{fig:xc100_dataset_summary}(d) partitions the data set into C/H-only species (12\%), carbon-containing species with additional elements (51\%), and carbon-free species (37\%). 

The collection includes species ranging from hydrocarbons such as benzene, cyclopropane, and cyclopentadiene to heterocycles such as furan, pyridine, and pyrrole, as well as biologically-derived molecules such as glycine and urea. Inorganic and main-group species include borazine (the B/N analogue of benzene), ammonia borane, boric acid, beryllium oxide, lithium borohydride, lithium nitrate, and nitrogen trifluoride. The nitrosonium and Zundel cations additionally extend the collection beyond neutral species.

\subsection{Quality of the Kohn-Sham Determinant Optimization}

\begin{figure}[H]
    \centering
    \includegraphics[width=0.78\textwidth]
    {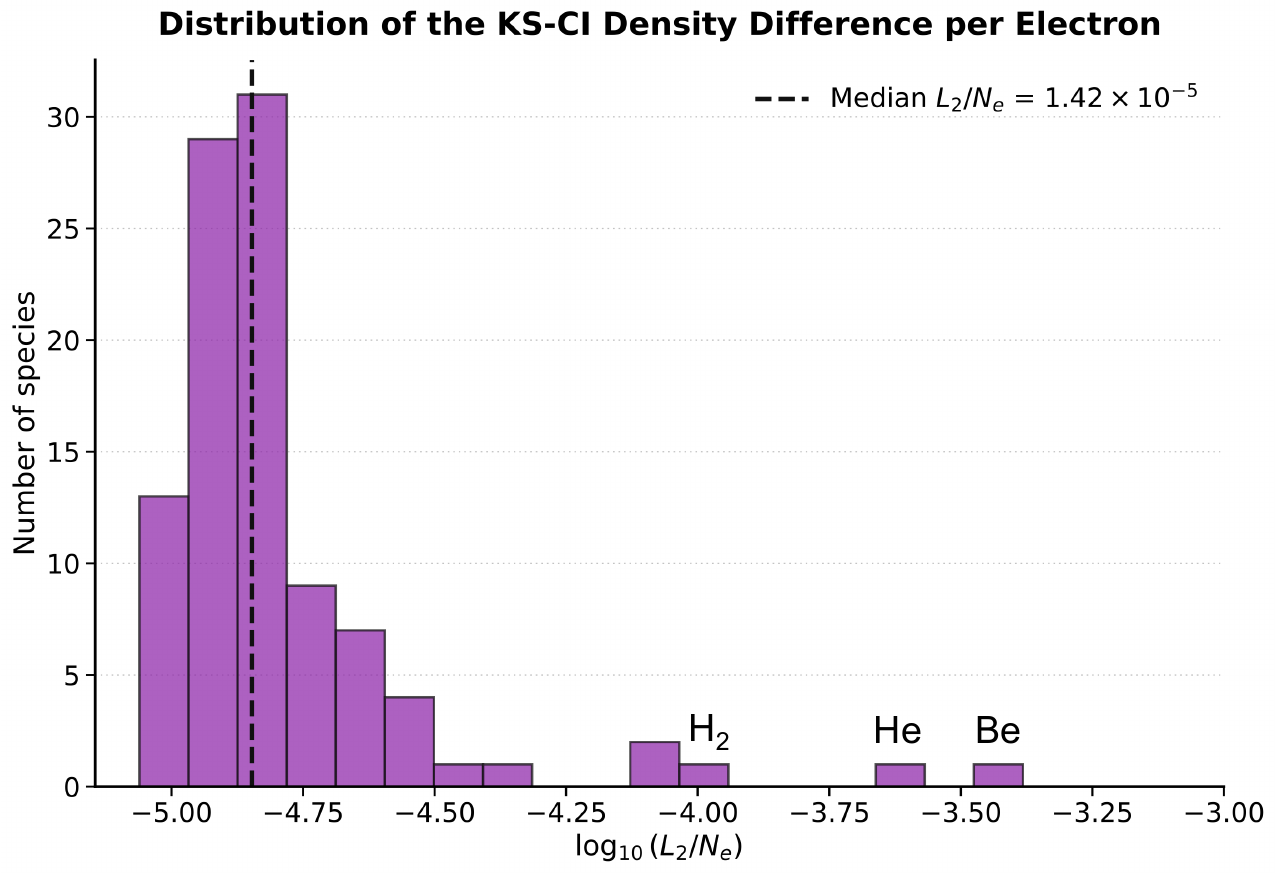}
    \caption{Distribution of the KS-CI density difference per electron across the XC100 data set. The density difference is quantified by the $L_2$ norm
    $\|\rho_{\mathrm{KS}}-\rho_{\mathrm{CI}}\|_2
    =
    [\int|\rho_{\mathrm{KS}}(\mathbf r)
    -\rho_{\mathrm{CI}}(\mathbf r)|^2\,d\mathbf r]^{1/2}$, and normalized by the number of electrons, $N_e$. The histogram is shown on the $\log_{10}(L_2/N_e)$ scale, and the dashed vertical line marks the median value. The prominent high-residual values corresponding to \ce{H2}, He and Be are labeled.}
    \label{fig:l2_per_electron_distribution}
\end{figure}

A central requirement in constructing XC100 is that the optimized non-interacting KS determinant closely reproduces the corresponding correlated WFT density. Figure~\ref{fig:l2_per_electron_distribution} summarizes the residual density differences across all 100 species using the $L_2$ norm divided by the number of electrons.

The distribution is concentrated at small density differences, with a median $L_2/N_e$ of $1.42\times10^{-5}$. Most species lie within approximately one order of magnitude of this value, with two species sitting outside this range. The highest error is for the Be atom, which has signatures of strong correlation via 2s-2p mixing in its wavefunction.\cite{Hait2021}

\begin{figure}[H]
    \centering
    \includegraphics[width=\textwidth]
    {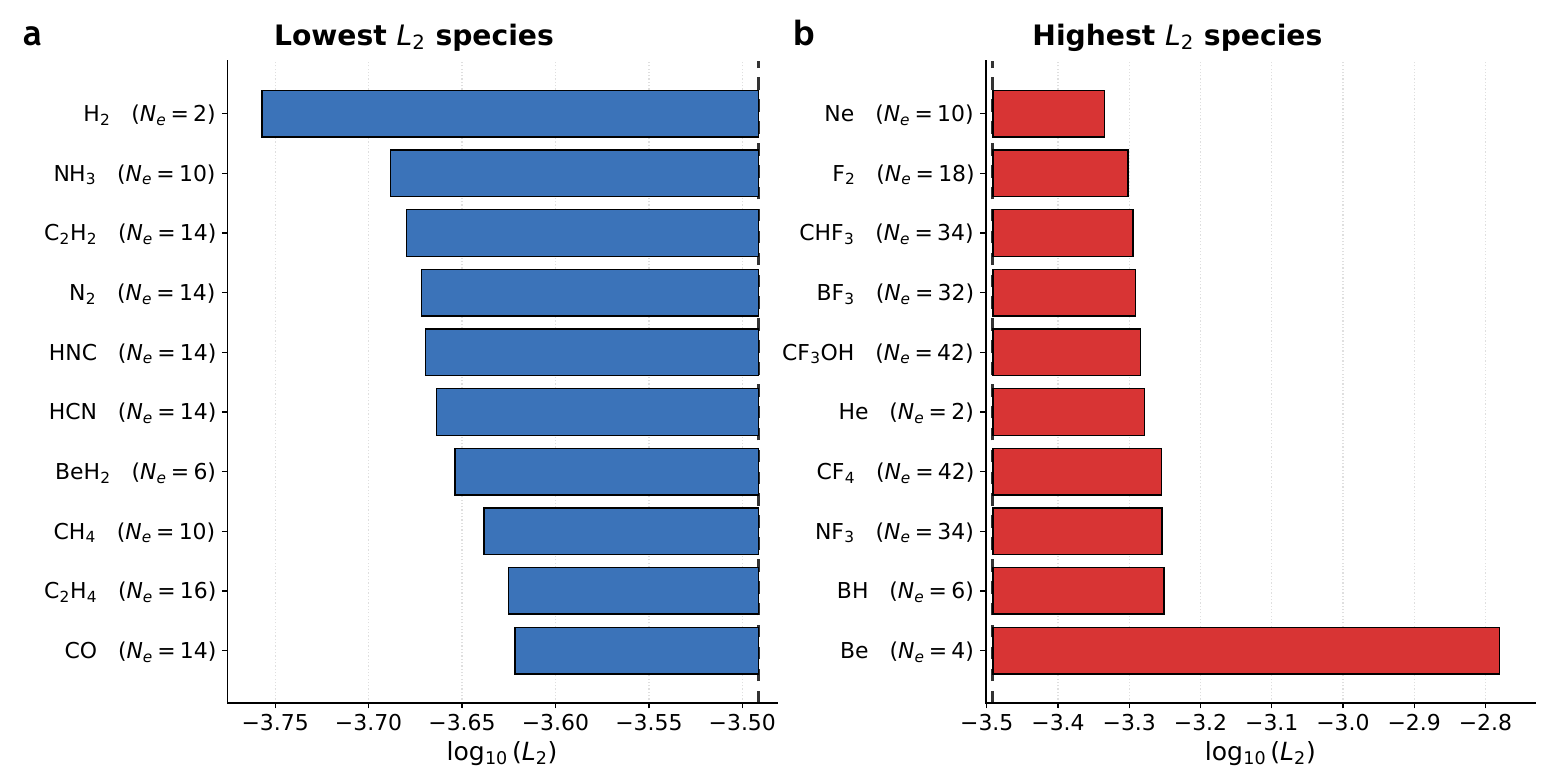}
    \caption{Species-resolved extremes of the absolute density-matching
    residual in XC100. The ten species with the smallest raw $L_2$
    residuals are shown in panel (a), and the ten species with the
    largest raw $L_2$ residuals are shown in panel (b). Bars extend from
    the global XC100 median,
    $L_2=3.22\times10^{-4}$, indicated by the dashed vertical line, to
    the value for each species. Electron counts are given in
    parentheses. The panels use different horizontal ranges to resolve
    the comparatively narrow low-residual and broader high-residual
    tails.}
    \label{fig:l2_species_extremes}
\end{figure}

For comparison to Figure~\ref{fig:l2_per_electron_distribution}, Figure~\ref{fig:l2_species_extremes} shows the extreme values for the \emph{unnormalized} $L_2$ metric. This $L_2$ data clarifies the interpretation of several low-electron-count species that appear in the high-residual tail after normalization. For example, \ce{H2} lies in the high-residual tail of the normalized distribution, with $L_2/N_e=8.75\times10^{-5}$, yet it has the smallest raw $L_2$ residual in XC100. In contrast, Be remains the clearest high-residual species even without normalization. The upper raw-$L_2$ tail also contains atoms and diatomics, such as He and BH, and small polyatomic species, including \ce{NF3}, \ce{CF4}, \ce{BF3}, and \ce{CF3OH}. Most of the polyatomic species remain in the low-residual region under both measures. Thus, the largest absolute residuals are not determined simply by electron count, and the raw and normalized metrics provide complementary views of the quality of the determinant optimization.

The magnitude of the density residuals can be placed in the context of previous WFT-to-KS calculations. SlaterRKS calculations in the QZ4P basis reported raw $L_2$ density differences ranging from approximately $1.5\times10^{-3}$ to $1.9\times10^{-2}$ for representative molecular systems.\cite{Tribedi2023} These values are somewhat larger than the XC100 median raw residual of $3.22\times10^{-4}$. In the finite-element inverse-DFT calculations of Kanungo \textit{et al.}, the $L_2$ density differences were just below $1\times10^{-4}$,\cite{Kanungo2019} a factor of about 4 below the XC100 median. For comparison with the original RKS method,\cite{Ryabinkin2015} we also computed $L_1$ density differences ($\|\rho_{\mathrm{KS}}-\rho_{\mathrm{CI}}\|_1=\int|\rho_{\mathrm{KS}}(\mathbf r)-\rho_{\mathrm{CI}}(\mathbf r)|\,d\mathbf r$) for two like-basis atomic cases. For He in the cc-pVTZ basis, the present KS determinant optimization gives $L_1=2.23\times10^{-3}$, compared with the reported RKS value of $2.51\times10^{-3}$. For Be in the cc-pCVTZ basis, we obtain $L_1=4.02\times10^{-3}$, while RKS gives $4.93\times10^{-3}$ using a CAS(2,4) wavefunction. The two methods therefore give closely comparable density residuals in these cases, with the present optimization yielding slightly smaller $L_1$ values.

Taken together, the median $L_2/N_e=1.42\times10^{-5}$ and median raw $L_2=3.22\times10^{-4}$ show that the inverted KS determinants closely reproduce the correlated target densities across XC100. The cc-pVTZ exact-exchange and kinetic-correlation quantities obtained from the KS determinant optimization are reported in Supporting Information Figure~S1, with their basis-set sensitivity examined in Supporting Information Table~S1.

The residual density difference characterizes the quality of the finite-basis WFT-to-KS mapping, but it is not itself a measure of the error in the composite $E_{xc}$. The validity of the composite approach will therefore also be tested in the next subsection by comparing finite-basis composite estimates with full cc-pCVQZ results.

\subsection{Magnitude and Components of the Composite Corrections}

The composite terms entering Eq. \ref{eq:composite_exc} are intended to recover correlation effects that remain outside the cc-pVTZ baseline. It is useful to establish the magnitude of these corrections, determine how the individual larger-basis contributions are distributed, and test whether retaining the cc-pVTZ KS mapping limits the final $E_{xc}$.

\begin{figure}[H]
    \centering
    \includegraphics[width=\textwidth]
    {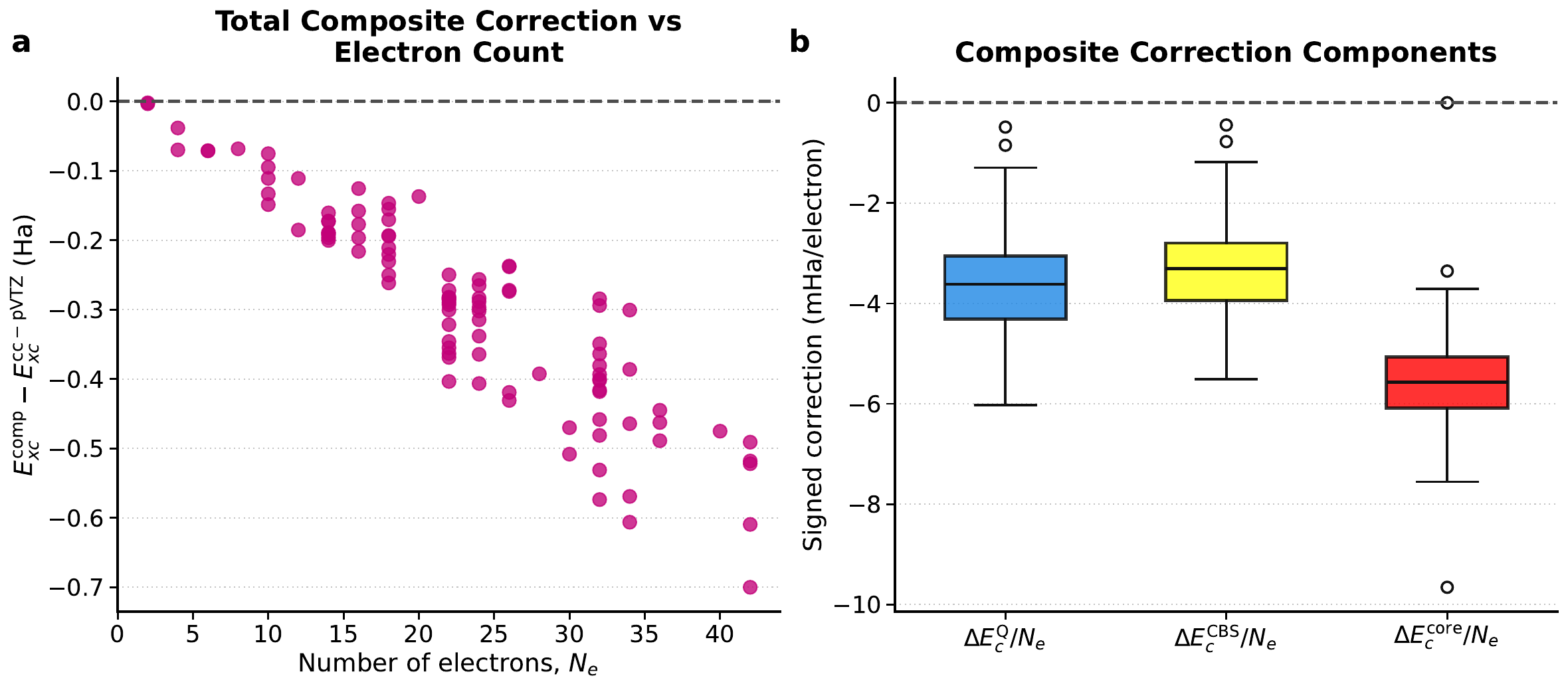}
    \caption{Magnitude and composition of the additive correlation-energy
    corrections used to construct the composite XC100 reference energies.
    (a) Total change from the cc-pVTZ XC energy to the final composite XC
    energy,
    $E_{xc}^{\mathrm{comp}}
    -E_{xc}^{\mathrm{cc\text{-}pVTZ}}$, as a function of electron count.
    The horizontal dashed line indicates zero correction.
    (b) Distributions of the cc-pVTZ-to-cc-pVQZ correction,
    $\Delta E_c^{\mathrm{Q}}$, the cc-pVQZ-to-CBS
    Riemann correction, $\Delta E_c^{\mathrm{CBS}}$, and the
    cc-pVTZ-to-cc-pCVTZ correction,
    $\Delta E_c^{\mathrm{core}}$, normalized by the number
    of electrons, $N_e$.
    In each box plot, the box spans the interquartile range, the
    horizontal line marks the median, and the whiskers extend to the
    most extreme values within $1.5$ times the interquartile range.
    Open circles denote values outside the whisker range.}
    \label{fig:composite_corrections}
\end{figure}

The magnitude and components of the composite corrections across XC100 are shown in Figure~\ref{fig:composite_corrections}. Panel (a) reports the total change $E_{xc}^{\mathrm{comp}}-E_{xc}^{\mathrm{cc\text{-}pVTZ}}$ as a function of electron count, whereas panel (b) separates the three correlation-energy corrections and normalizes each by $N_e$.

The composite correction generally increases in magnitude with electron count [Figure~\ref{fig:composite_corrections}(a)]. Thus, the larger-basis correlation treatment systematically lowers $E_{\mathrm{xc}}$ relative to the cc-pVTZ reference. The broad increase in magnitude with $N_e$ is consistent with the extensive nature of the correlation energy: as the number of electrons increases, a larger absolute amount of correlation energy remains to be recovered. At the same time, species with similar electron counts can exhibit appreciably different corrections, indicating that the basis-set differences are not determined by electron count alone.

Normalizing the individual terms ($\Delta E_c^{\mathrm{Q}}$, $\Delta E_c^{\mathrm{CBS}}$ and $\Delta E_c^{\mathrm{core}}$) by electron count makes their relative contributions more apparent [Figure~\ref{fig:composite_corrections}(b)]. All three corrections are typically on the scale of several mHa per electron. The TZ-to-QZ $\Delta E_c^{\mathrm{Q}}$ and QZ-to-CBS Riemann $\Delta E_c^{\mathrm{CBS}}$ terms are closely related because, for the two-point extrapolation used here, $\Delta E_c^{\mathrm{CBS}}=A_4\Delta E_c^{\mathrm{Q}}$, with $A_4\approx0.914$\cite{Riemann}. The Riemann term therefore contributes a CBS tail that is comparable to, but slightly smaller in magnitude than, the TZ-to-QZ correction. The cc-pCVTZ $\Delta E_c^{\mathrm{core}}$ correction shows the largest median magnitude of the three components, demonstrating the additional flexibility of the core-valence-optimized basis. Regardless, all three terms have similar magnitude and none can be neglected in the total composite energy.

An important question remains: does the KS mapping at the cc-pVTZ level limit the accuracy of the resulting composite $E_{xc}$? This question is partly motivated by the observation that the individual KS-derived quantities are not converged with respect to the orbital basis. In particular, full cc-pCVQZ inversions for \ce{BH3}, \ce{CH4}, and \ce{NH3} show changes in $T_c$ of approximately $4.0$-$4.9$ mHa per electron relative to cc-pVTZ. $E_x$, however, is well converged with the corresponding changes of at most approximately $0.65$ mHa per electron (Supporting Information, Table~S1).

To test whether this basis dependence propagates into the composite $E_{xc}$, inverse calculations with the cc-pCVQZ basis were performed for the three species listed above. For this comparison, the Riemann cc-pVQZ-to-CBS contribution was omitted from the composite construction so that the composite estimate and the direct reference both represent finite-basis quantities. The results are summarized in Table~\ref{tab:composite_validation}.

\begin{table}[H]
\centering
\small
\setlength{\tabcolsep}{4pt}
\caption{Basis-set sensitivity of the kinetic-correlation energy and
the composite $E_{xc}$ construction.
$T_c^{\mathrm{TZ}}$ and $T_c^{\mathrm{CVQZ}}$ are obtained from
inverse calculations with the cc-pVTZ and cc-pCVQZ basis sets,
respectively. $|\Delta T_c|/N_e$ is their absolute difference per
electron. $E_{xc}^{\mathrm{comp,no\,CBS}}$ is the composite estimate
with the Riemann CBS contribution omitted, and
$|\Delta E_{xc}|/N_e$ is its absolute difference from the full
cc-pCVQZ KS determinant optimization.}
\label{tab:composite_validation}
\begin{tabular}{lcccccc}
\hline
Species &
$T_c^{\mathrm{TZ}}$ &
$T_c^{\mathrm{CVQZ}}$ &
$|\Delta T_c|/N_e$ &
$E_{xc}^{\mathrm{CVQZ}}$ &
$E_{xc}^{\mathrm{comp,no\,CBS}}$ &
$|\Delta E_{xc}|/N_e$ \\
&
(Ha) &
(Ha) &
(mHa/electron) &
(Ha) &
(Ha) &
(mHa/electron) \\
\hline
\ce{BH3} &
0.122918 &
0.161970 &
4.881 &
-5.124182 &
-5.126027 &
0.230 \\

\ce{CH4} &
0.186176 &
0.225974 &
3.979 &
-6.884075 &
-6.885586 &
0.151 \\

\ce{NH3} &
0.205119 &
0.252812 &
4.769 &
-8.000418 &
-8.010314 &
0.989 \\
\hline
\end{tabular}
\end{table}

The finite-basis composite $E_{xc}$ values reproduce the corresponding full cc-pCVQZ inversions to within 1 mHa per electron for the three species examined. Thus, convergence of the individual KS-derived components is not required for the composite construction to recover the larger-basis exchange-correlation energy accurately. This is consistent with the design of the approach: the cc-pVTZ basis supplies the KS reference, while the larger-basis correction is obtained from changes in the WFT correlation energy rather than from separate extrapolation of $T_s$, $T_c$, or $E_x$.

\subsection{DFA Errors Relative to the XC100 Reference Energies}

Most DFA benchmarks assess total or relative energies, for which errors in the individual terms of the KS energy decomposition can partially cancel.\cite{Goerigk2017,Gordon_Exc,Liang2025} XC100 instead provides a WFT-derived reference for $E_{\mathrm{xc}}$, allowing the exchange-correlation contribution itself, the quantity approximated in practical KS-DFT, to be assessed. 
This provides a complementary test to conventional energetic benchmarking by asking how accurately a DFA reproduces the XC contribution.

\begin{figure}[H]
    \centering
    \includegraphics[width=0.80\textwidth]{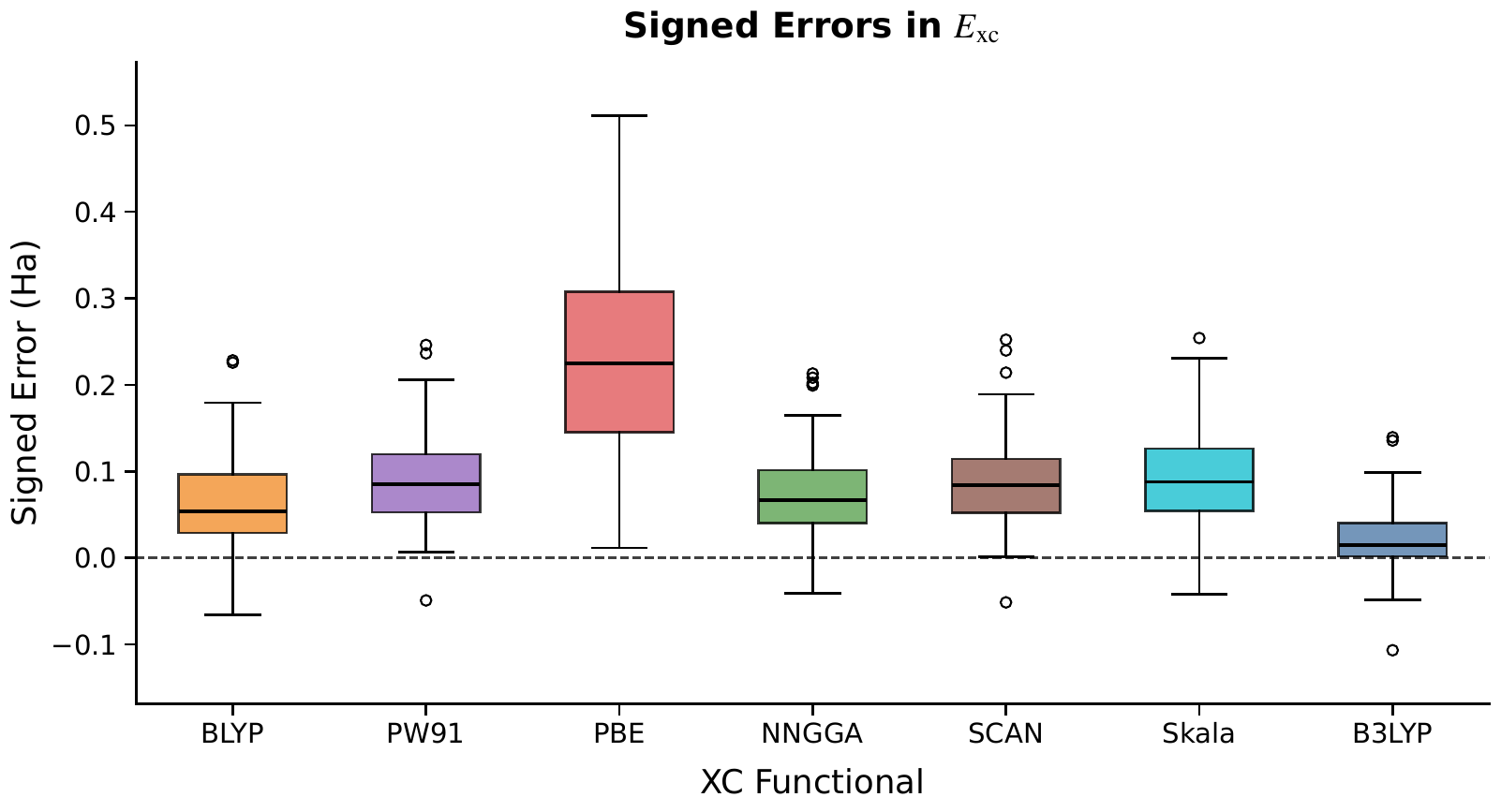}
    \caption{Distribution of signed errors in the exchange-correlation energy, $E_{\mathrm{xc}}$, for selected density-functional approximations, relative to the composite XC100 reference energies. Positive values indicate that the DFA value is less negative than the reference value. The conventional functionals include all 100 XC100 species. The NNGGA and Skala distributions contain the 95 and 97 species, respectively, for which the self-consistent calculations converged.} 
    \label{fig:exc_signed_errors}
\end{figure}

Figure~\ref{fig:exc_signed_errors} compares the composite XC100 $E_{\mathrm{xc}}$ references with values obtained from conventional and machine-learned density-functional approximations. The signed error is defined as the DFA value minus the XC100 reference value, so a positive error indicates a DFA $E_{\mathrm{xc}}$ that is less negative than the WFT-derived reference. All DFA results are from fully self-consistent computations.

The conventional semilocal functionals exhibit predominantly positive $E_{\mathrm{xc}}$ errors. PBE\cite{PBE_1,PBE_2} shows the largest shift, with its median well above zero and the largest spread in errors. BLYP\cite{B88,LYP_1,LYP_2}, PW91\cite{PW91_1,PW91_2}, and SCAN\cite{SCAN} show smaller deviations, while B3LYP's\cite{B3LYP} error profile is centered closest to zero and has a substantially narrower interquartile range. The two machine-learned functionals NNGGA\cite{kanungo2025} and Skala\cite{Luise2026} also exhibit predominantly positive signed errors. NNGGA, which is constructed as a neural-network correction to a PBE baseline, substantially reduces the positive shift observed for PBE and yields a comparatively compact $E_{\mathrm{xc}}$ error distribution. The Skala distribution has a larger positive median and spread than NNGGA. 

The NNGGA result is particularly relevant to the motivation behind XC100. Unlike most functionals trained primarily against total or relative energies, the NNGGA training included exact $E_{\mathrm{xc}}$ values together with density-weighted $v_{\mathrm{xc}}$ obtained from inverse DFT.\cite{kanungo2025} The functional was trained using only a small set of atoms and molecules, yet Kanungo \textit{et al.} showed that the resulting GGA attains thermochemical accuracy comparable to the higher-rung SCAN meta-GGA.\cite{kanungo2025} The relatively small $E_{\mathrm{xc}}$ errors across the XC100 set are therefore consistent with the value of using $E_{\mathrm{xc}}$ targets in functional training.

Skala provides a complementary comparison because it was developed using a different training strategy. Its hundreds of thousands of high-accuracy training labels consist of wavefunction-level energy differences spanning atomization energies, reaction energetics, ionization and proton affinities, conformational energies, and related chemical data, but no reference $E_{\mathrm{xc}}$ values.\cite{Luise2026} Skala achieves high accuracy across these total- and relative-energy benchmarks, while Figure~\ref{fig:exc_signed_errors} shows that its absolute $E_{\mathrm{xc}}$ errors on XC100 remain appreciable. This further illustrates that XC100 probes a component of the KS energy that is not directly isolated by conventional energetic benchmarks.


The interested reader can find in the Supporting Information Figure~\ref{fig:ex_ec_signed_errors} additional comparisons of DFAs with respect to XC100's benchmark $E_x$ and $E_c$ values. While most DFAs are not designed to provide either term on their own---being constructed to model the total $E_xc$---this information is available in the XC100 dataset nonetheless. In the future, it may be possible to use the $E_c$ values from the XC100 workflow to design functionals based on 100\% exact exchange, providing more motivation to factor the XC energies into these terms.



\section{Conclusions}
This work introduced XC100, a data set of wavefunction-derived exchange-correlation energies for 100 closed-shell atomic and molecular species. Accurate CI wavefunctions provide the reference energies and densities, while KS inversion supplies the non-interacting kinetic energy needed to construct $E_{\mathrm{xc}}$. Across XC100, the resulting KS densities closely reproduce their correlated wavefunction counterparts, with a median $L_2/N_e$ difference of $1.42\times10^{-5}$. The procedure therefore provides a practical route for obtaining KS-resolved reference information from high-accuracy wavefunction calculations over a substantially broader chemical space than has typically been accessible to inverse WFT-to-KS approaches.\cite{shi_wasserman_2021,Kanungo2019,Tribedi2023,Khanna2026}

The composite scheme introduced accurately recovers larger-basis correlation, core-valence basis-set effects, and the remaining correlation-energy contribution toward the complete-basis-set limit without repeating the KS determinant optimization at each basis level. For \ce{BH3}, \ce{CH4}, and \ce{NH3}, the finite-basis composite values agree with full cc-pCVQZ KS determinant optimizations to within 1 mHa per electron. Together with the Riemann extrapolation, these corrections bring the XC100 reference energies toward the nonrelativistic complete-basis-set limit while retaining a computationally tractable cc-pVTZ KS mapping.


XC100 provides direct reference data for $E_{\mathrm{xc}}$ itself, complementing conventional benchmarks based primarily on total and relative energies. The DFA comparisons presented here illustrate that such XC benchmarks can reveal information not apparent from total or relative energetic benchmarks alone. More broadly, the workflow is not restricted to the closed-shell systems considered here and can be extended to larger, open-shell, and strongly correlated species as suitable wavefunction energies and densities become available. Such extensions would broaden the range of XC benchmarks and training targets available for future density-functional development.


\section*{Acknowledgements}

This project has been supported by the U.S. Department of Energy through the grant DE-SC0022241. This research used resources of the National Energy Research Scientific Computing Center (NERSC), a U.S. Department of Energy User Facility using NERSC awards BES-ERCAP0034270 and BES-ERCAP0037081.

\section*{Supporting information}

The following file is available free of charge.

\begin{itemize}

    \item \texttt{XC100\_Supporting\_Information.pdf}: Supporting Information containing additional computational details for the reference wavefunction, KS determinant optimization, and DFT calculations; numerical sensitivity tests for the reference calculations and composite correction scheme; additional analysis of the reference exchange and correlation quantities; separate DFA exchange- and correlation-energy error distributions, and machine-learned functional SCF convergence details.

\end{itemize}

\section*{Conflicts of Interest}
There are no conflicts to declare.

\section*{Author Contributions}
Vaibhav Khanna: Conceptualization (equal); Data curation (equal); Formal analysis (equal); Investigation (equal); Methodology (equal); Software (equal); Visualization
(equal); Writing – original draft (equal); Writing – review \& editing (equal). Paul M. Zimmerman: Conceptualization (equal); Formal analysis (equal); Funding acquisition (equal); Investigation (equal); Project administration (equal); Software (equal); Supervision (equal); Writing – review \& editing (equal).

\section*{Data Availability}
The XC100 data set, Cartesian structures, molecule-to-system-ID mapping, and CI-derived Kohn-Sham one-particle reduced density matrices (1-RDMs) in the cc-pVTZ AO basis are available from the Zimmerman Group GitHub repository at \url{https://github.com/ZimmermanGroup/XC100}.




\clearpage

\printbibliography


\clearpage

\setcounter{figure}{0}
\setcounter{table}{0}
\setcounter{equation}{0}
\setcounter{section}{0}

\renewcommand{\thefigure}{S\arabic{figure}}
\renewcommand{\thetable}{S\arabic{table}}
\renewcommand{\theequation}{S\arabic{equation}}
\renewcommand{\thesection}{S\arabic{section}}
\renewcommand{\thesubsection}{S\arabic{section}.\arabic{subsection}}

\clearpage

\begin{center}
    {\Large\bfseries Supporting Information}\\[1em]
    {\large XC100: A Wavefunction-Derived Exchange-Correlation Energy
    Dataset for Atomic and Molecular Species}\\[1em]
    Vaibhav Khanna and Paul M. Zimmerman*
\end{center}

\vspace{1em}

\startcontents[si]

\section*{Contents}
\printcontents[si]{}{1}{}

\section{Additional Computational Details}

Reference wavefunctions were obtained directly with HBCI for all atoms and diatomics, while the remaining molecular species were treated using iFCI with HBCI as the solver for the individual increments. The direct HBCI calculations for atoms and diatomics were performed in a natural-orbital (NO) basis. An initial HBCI calculation was used to construct the one-particle density matrix, whose eigenvectors defined the NOs used in the final HBCI calculation. Starting from a complete-active-space configuration-interaction reference containing up to eight electrons in eight orbitals, the HBCI variational space was constructed iteratively by selecting determinants according to the $\varepsilon_1$ threshold. The perturbative correction was then evaluated using the $\varepsilon_2$ threshold. \cite{holmes2016,sharma2017,li2018,dang2023,chien2018}

For the molecular species treated using iFCI, the perfect-pairing (PP) procedure began with Pipek-Mezey localization of the Hartree-Fock orbitals, followed by construction of the initial virtual orbitals using the Sano procedure and full orbital optimization under the pairing ansatz. \cite{Small2012,Pipek1989,Lawler2010,Janesko2022,VanVoorhis2001,Cullen1996,Cooper2007,Boys1960} An HBCI solver was used to evaluate the correlation energy associated with each iFCI increment.

For each increment, the selected occupied orbitals were correlated with a virtual space constructed from natural orbitals. The NOs were generated from an approximate CI calculation and screened according to their occupation numbers to reduce the dimensions of the individual CI calculations.\cite{Hatch2025} The screening parameter $\zeta$ defines the corresponding occupation-number cutoff: a virtual NO with an occupation number below $10^{-\zeta}$ is excluded from the correlated orbital space. Increasing $\zeta$ therefore lowers the occupation-number cutoff and retains a larger fraction of the virtual space. The resulting active-space CI problem for each increment was then solved using HBCI. For the one- and two-body increments, $\zeta=8$ was used, corresponding to exclusion of virtual NOs with occupation numbers below $10^{-8}$. For the three-body increments, $\zeta=5.5$ was used, corresponding to exclusion of virtual NOs with occupation numbers below $10^{-5.5}$. The reduced virtual space at the three-body level lowers the cost of these substantially larger CI calculations. The numerical sensitivity of the three-body energies to this choice is reported in Table~\ref{tab:zeta_sensitivity}.

\subsection{Kohn-Sham Determinant Optimization}

The KS determinant optimization was performed using an L-BFGS quasi-Newton algorithm. At each optimization iteration, a collective search direction was constructed from the complete occupied-virtual orbital-rotation gradient together with a limited history of previous steps and gradient changes. The orbital-rotation parameters were assembled into an antisymmetric matrix $\boldsymbol{\kappa}$ and applied as

\begin{equation}
    \mathbf{C}_{\mathrm{new}}
    =
    \mathbf{C}\exp(\boldsymbol{\kappa}),
\end{equation}

which preserves the orthonormality condition

\begin{equation}
    \mathbf{C}^{T}\mathbf{S}\mathbf{C}
    =
    \mathbf{I}
\end{equation}

throughout the optimization. A line search over progressively shorter trial steps was used to select the step length. Convergence was reached when the Euclidean norm of the complete orbital-rotation gradient was below $10^{-7}$.

The weighting parameter $\lambda$ entering the KS determinant optimization objective in Eq.~\ref{eq:ks_orbital_generation} of the main text was set to $\lambda=1\times10^{-5}$ for the XC100 calculations. Its numerical influence on the derived $E_{\mathrm{xc}}$ values is examined in Table~\ref{tab:lambda_sensitivity}, while the basis-set dependence of the KS-derived $E_x$ and $T_c$ quantities is examined in Table~\ref{tab:basis_sensitivity}.

\subsection{Evaluation of Density-Functional Approximation Energies}

All conventional DFA calculations used a PySCF\cite{PySCF} level-9 numerical grid with 200 radial and 1454 angular points, with grid pruning disabled. For each semilocal functional, the exchange-correlation, exchange, and correlation energies were evaluated separately by numerical integration over the converged self-consistent density, and the decomposition

\begin{equation}
    E_{\mathrm{xc}}
    =
    E_{\mathrm{x}}
    +
    E_{\mathrm{c}}
\end{equation}

was verified numerically.

For B3LYP, the exchange-correlation energy was constructed explicitly as

\begin{equation}
E_{\mathrm{xc}}^{\mathrm{B3LYP}}
=
0.20E_{\mathrm{x}}^{\mathrm{HF}}
+
0.08E_{\mathrm{x}}^{\mathrm{Slater}}
+
0.72E_{\mathrm{x}}^{\mathrm{B88}}
+
0.81E_{\mathrm{c}}^{\mathrm{LYP}}
+
0.19E_{\mathrm{c}}^{\mathrm{VWN\text{-}RPA}}.
\label{eq:si_b3lyp_definition}
\end{equation}
\cite{B3LYP,Becke_Hybrid,VWN}

The semilocal exchange and correlation contributions were evaluated by numerical quadrature over the converged B3LYP density. The unscaled Hartree-Fock exact-exchange contribution was evaluated from the closed-shell exchange matrix as

\begin{equation}
E_{\mathrm{x}}^{\mathrm{HF}}
=
-\frac{1}{4}
\operatorname{Tr}
\left[
\mathbf{P}\mathbf{K}
\right],
\label{eq:si_b3lyp_hf_exchange}
\end{equation}

where $\mathbf{P}$ is the spin-summed restricted Kohn-Sham density matrix and $\mathbf{K}$ is the corresponding exchange matrix. The scaled exact-exchange contribution was then combined with the semilocal terms in Eq.~\ref{eq:si_b3lyp_definition} to obtain the reported B3LYP $E_{\mathrm{xc}}$.

\section{HBCI Composite-Correction Expressions}

For HBCI wavefunctions, the correlation energy in basis $B$ is defined relative to the corresponding Hartree-Fock energy as

\begin{equation}
E_{c,\mathrm{HBCI}}^{B}
=
E_{\mathrm{HBCI}}^{B}
-
E_{\mathrm{HF}}^{B}.
\end{equation}

The finite-basis corrections entering the composite energy are

\begin{equation}
\Delta E_{c}^{\mathrm{cc\text{-}pVQZ}}
=
E_{c,\mathrm{HBCI}}^{\mathrm{cc\text{-}pVQZ}}
-
E_{c,\mathrm{HBCI}}^{\mathrm{cc\text{-}pVTZ}},
\end{equation}

and

\begin{equation}
\Delta E_{c}^{\mathrm{cc\text{-}pCVTZ}}
=
E_{c,\mathrm{HBCI}}^{\mathrm{cc\text{-}pCVTZ}}
-
E_{c,\mathrm{HBCI}}^{\mathrm{cc\text{-}pVTZ}}.
\end{equation}

The corresponding two-point Riemann CBS estimate is

\begin{equation}
E_{c,\mathrm{HBCI}}^{\mathrm{CBS}(T,Q)}
=
E_{c,\mathrm{HBCI}}^{\mathrm{cc\text{-}pVQZ}}
+
A_4
\left(
E_{c,\mathrm{HBCI}}^{\mathrm{cc\text{-}pVQZ}}
-
E_{c,\mathrm{HBCI}}^{\mathrm{cc\text{-}pVTZ}}
\right),
\end{equation}

with the additional cc-pVQZ-to-CBS correction

\begin{equation}
\Delta E_{c}^{\mathrm{Riemann}}
=
E_{c,\mathrm{HBCI}}^{\mathrm{CBS}(T,Q)}
-
E_{c,\mathrm{HBCI}}^{\mathrm{cc\text{-}pVQZ}}.
\end{equation}


\section{Numerical Sensitivity of the Reference Calculations}

To assess the sensitivity of the XC100 reference quantities to selected computational choices, additional calculations were performed for \ce{BH3}, \ce{CH4}, and \ce{NH3}. These tests examine the basis-set dependence of the iFCI 3-body correlation energies, KS-derived exact-exchange and kinetic-correlation energies, the dependence of $E_{\mathrm{xc}}$ on the KS kinetic energy weighting parameter $\lambda$, and the sensitivity of the iFCI energies to the $\epsilon_1$ and $\zeta$ parameters.

\subsection{Basis-Set Sensitivity of $E_{\mathrm{x}}$ and $T_{\mathrm{c}}$}

Table~\ref{tab:basis_sensitivity} compares the exact-exchange and kinetic-correlation energies obtained with the cc-pVTZ and cc-pCVQZ basis sets for \ce{BH3}, \ce{CH4}, and \ce{NH3}.

\begin{table}[H]
    \centering
    \caption{Basis-set dependence of the exact-exchange energy,
    $E_{\mathrm{x}}$, and kinetic-correlation energy,
    $T_{\mathrm{c}}$, for \ce{BH3}, \ce{CH4}, and \ce{NH3}.
    Energies are reported in Hartree. The per-electron changes are
    defined as
    $\Delta E_{\mathrm{x}}/N_e =
    (E_{\mathrm{x}}^{\mathrm{cc\text{-}pCVQZ}}
    -E_{\mathrm{x}}^{\mathrm{cc\text{-}pVTZ}})/N_e$
    and
    $\Delta T_{\mathrm{c}}/N_e =
    (T_{\mathrm{c}}^{\mathrm{cc\text{-}pCVQZ}}
    -T_{\mathrm{c}}^{\mathrm{cc\text{-}pVTZ}})/N_e$. The corresponding full $E_{\mathrm{xc}}$ values are included to provide the scale of the basis-set changes in the individual KS-derived components.}
    \label{tab:basis_sensitivity}

    \small
    \resizebox{\textwidth}{!}{%
    \begin{tabular}{lrrrrrrll}
        \hline
        Species &
        $E_{\mathrm{x}}$ (cc-pCVQZ) &
        $E_{\mathrm{x}}$ (cc-pVTZ) &
        $\Delta E_{\mathrm{x}}/N_e$ &
        $T_{\mathrm{c}}$ (cc-pCVQZ) &
        $T_{\mathrm{c}}$ (cc-pVTZ) &
        $\Delta T_{\mathrm{c}}/N_e$  & $E_{\mathrm{xc}}$ (cc-pCVQZ)&$E_{\mathrm{xc}}$ (cc-pVTZ)\\
        \hline

        \ce{BH3} &
        -4.927391 &
        -4.922224 &
         -0.000645&
         0.161970 &
         0.122918 &
        0.004881 & -5.124182 &-5.075683 \\

        \ce{CH4} &
        -6.592484 &
        -6.586348 &
         -0.000613&
         0.225974 &
         0.186176 &
        0.003979 & -6.884075 &-6.830308 \\

        \ce{NH3} &
        -7.669745 &
        -7.670128 &
        0.000038&
         0.252812 &
         0.205119 &
        0.004769 & -8.000418 &-7.942591 \\

        \hline
    \end{tabular}%
    }
\end{table}

The exact-exchange energy shows relatively weak basis-set dependence for the three species examined. The change in $E_{\mathrm{x}}/N_e$ between cc-pVTZ and cc-pCVQZ is at most approximately $0.65$ mHa per electron. In comparison, $T_{\mathrm{c}}$ exhibits a larger basis-set dependence, with per-electron changes of approximately $4.0$-$4.9$ mHa. Thus, the kinetic-correlation energy is more sensitive to the basis-set size than the exact-exchange energy.


\subsection{Basis-Set Dependence of the Three-Body iFCI Correlation Energy}

The cc-pVTZ reference calculations include the three-body iFCI contribution, whereas the cc-pVTZ-to-cc-pVQZ composite correction is constructed from the changes in the one- and two-body iFCI contributions. To assess whether an additional basis-set correction to the three-body contribution is necessary, the three-body correlation energy was compared directly between the cc-pVTZ and cc-pVQZ basis sets for \ce{BH3}, \ce{CH4}, and \ce{NH3}.

\begin{table}[H]
    \centering
    \caption{Basis-set dependence of the three-body iFCI correlation
    energy, $E_{\mathrm{3c}}$, for \ce{BH3}, \ce{CH4}, and \ce{NH3}.
    The difference is defined as
    $\Delta E_{\mathrm{3c}}^{\mathrm{QZ-TZ}}
    =
    E_{\mathrm{3c}}^{\mathrm{cc\text{-}pVQZ}}
    -
    E_{\mathrm{3c}}^{\mathrm{cc\text{-}pVTZ}}$.
    All energies and energy differences are reported in Hartree.}
    \label{tab:three_body_basis_sensitivity}

    \small
    \begin{tabular}{lccc}
        \hline
        Species &
        $E_{\mathrm{3c}}^{\mathrm{cc\text{-}pVQZ}}$ &
        $E_{\mathrm{3c}}^{\mathrm{cc\text{-}pVTZ}}$ &
        $\Delta E_{\mathrm{3c}}^{\mathrm{QZ-TZ}}$ \\
        \hline

        \ce{BH3} &
        -0.001742 &
        -0.001722 &
        -0.000020 \\

        \ce{CH4} &
        -0.004012 &
        -0.003864 &
        -0.000148 \\

        \ce{NH3} &
        -0.002070 &
        -0.002102 &
         0.000032 \\

        \hline
    \end{tabular}
\end{table}

The three-body correlation contribution changes very little between cc-pVTZ and cc-pVQZ for the species examined. The absolute cc-pVTZ-to-cc-pVQZ differences are $0.020$, $0.148$, and $0.032$ mHa for \ce{BH3}, \ce{CH4}, and \ce{NH3}, respectively. After normalization by electron count, these correspond to only $2.5$, $14.8$, and $3.2$ $\mu$Ha per electron for \ce{BH3}, \ce{CH4}, and \ce{NH3}, respectively, with a maximum difference of $14.8$ $\mu$Ha per electron. Thus, the three-body correlation contribution is effectively unchanged upon increasing the basis from cc-pVTZ to cc-pVQZ for these representative species. This supports retaining the three-body contribution from the cc-pVTZ baseline while constructing the cc-pVTZ-to-cc-pVQZ composite correction from the basis-set changes in the one- and two-body contributions only.

\subsection{Sensitivity to the KS Determinant Optimization Parameter $\lambda$}

The KS determinant optimization parameter $\lambda$ in Eq.~\ref{eq:ks_orbital_generation} of the main text controls the relative weighting of the non-interacting kinetic-energy contribution in the optimization. Table~\ref{tab:lambda_sensitivity} compares the resulting cc-pVTZ exchange-correlation energies over an order-of-magnitude variation in $\lambda$.

\begin{table}[H]
    \centering
    \caption{Sensitivity of the cc-pVTZ exchange-correlation energy,
    $E_{\mathrm{xc}}$, to the KS determinant optimization weighting parameter
    $\lambda$. All energies are reported in Hartree.}
    \label{tab:lambda_sensitivity}

    \small
    \begin{tabular}{lccc}
        \hline
        Species &
        $E_{\mathrm{xc}}$ ($\lambda=1\times10^{-5}$) &
        $E_{\mathrm{xc}}$ ($\lambda=5\times10^{-5}$) &
        $E_{\mathrm{xc}}$ ($\lambda=1\times10^{-4}$) \\
        \hline

        \ce{BH3} &
        -5.075683 &
        -5.073950 &
        -5.072589 \\

        \ce{CH4} &
        -6.830308 &
        -6.828496 &
        -6.826793 \\

        \ce{NH3} &
        -7.942591 &
        -7.940929 &
        -7.939207 \\

        \hline
    \end{tabular}
\end{table}

The dependence of $E_{\mathrm{xc}}$ on $\lambda$ is small over the range examined. Increasing $\lambda$ by an order of magnitude, from $1\times10^{-5}$ to $1\times10^{-4}$, changes the total $E_{\mathrm{xc}}$ by only $3.1$-$3.5$ mHa for these species, corresponding to approximately $0.34$-$0.39$ mHa per electron. The change is also small relative to the magnitude of the corresponding $E_{\mathrm{xc}}$ values. Hence, the derived $E_{\mathrm{xc}}$ values are only weakly sensitive to $\lambda$ over the range considered here.

\subsection{Sensitivity of iFCI Energies to $\epsilon_1$ and $\zeta$}

The numerical stability of the iFCI energies was further examined by varying the $\epsilon_1$ and $\zeta$ parameters for \ce{BH3}, \ce{CH4}, and \ce{NH3}.

\begin{table}[H]
    \centering
    \caption{Sensitivity of the cc-pVQZ two-body iFCI energy to the
    $\epsilon_1$ parameter. All energies are reported in Hartree.}
    \label{tab:epsilon1_sensitivity}

    \small
    \begin{tabular}{lccc}
        \hline
        Species &
        $\epsilon_1=1\times10^{-4}$ &
        $\epsilon_1=2\times10^{-4}$ &
        $\epsilon_1=5\times10^{-4}$ \\
        \hline

        \ce{BH3} &
        -26.569863 &
        -26.569860 &
        -26.569869 \\

        \ce{CH4} &
        -40.475182 &
        -40.475155 &
        -40.475157 \\

        \ce{NH3} &
        -56.519505 &
        -56.519450 &
        -56.519396 \\

        \hline
    \end{tabular}
\end{table}

The cc-pVQZ two-body iFCI energies show very little dependence on $\epsilon_1$ over the tested range, indicating that the two-body iFCI energies are well converged with respect to $\epsilon_1$ at the precision relevant to the present calculations.

\begin{table}[H]
    \centering
    \caption{Sensitivity of the cc-pCVQZ three-body iFCI energy to
    the $\zeta$ parameter. The final column gives
    $\Delta E =
    E_{\mathrm{iFCI}}(\zeta=5.5)
    -E_{\mathrm{iFCI}}(\zeta=8)$.
    All energies and energy differences are reported in Hartree.}
    \label{tab:zeta_sensitivity}

    \small
    \begin{tabular}{lccc}
        \hline
        Species &
        $E_{\mathrm{iFCI}}^{(3)}$ ($\zeta=8$) &
        $E_{\mathrm{iFCI}}^{(3)}$ ($\zeta=5.5$) &
        $\Delta E$ \\
        \hline

        \ce{BH3} &
        -26.595824 &
        -26.595817 &
         0.000007 \\

        \ce{CH4} &
        -40.503115 &
        -40.503122 &
        -0.000007 \\

        \ce{NH3} &
        -56.548773 &
        -56.548754 &
         0.000019 \\

        \hline
    \end{tabular}
\end{table}

Changing $\zeta$ from 8 to 5.5 changes the cc-pCVQZ three-body iFCI energies by only $0.007$--$0.019$ mHa for the three species examined. This weak sensitivity reflects the efficiency of the iNO procedure:\cite{Hatch2025} virtual orbitals are ranked by their natural occupations, so tightening $\zeta$ from 5.5 to 8 primarily restores very weakly occupied virtual NOs that contribute negligibly to the correlation energy. The iNO screening therefore provides a substantial reduction in the three-body virtual space while retaining the energetically important correlation contributions.

\section{Characteristics of the Reference Exchange and Correlation Quantities}

\begin{figure}[H]
    \centering
    \includegraphics[width=\textwidth]
    {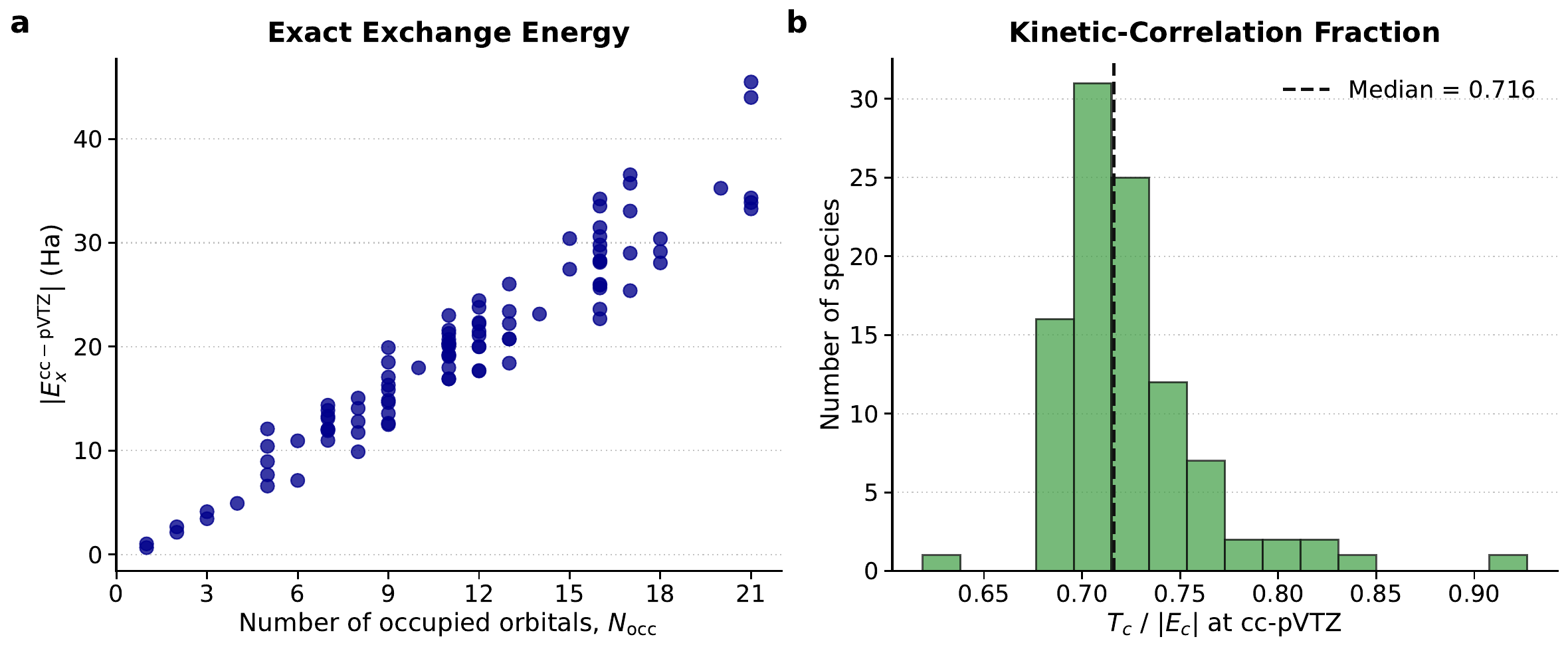}
    \caption{Reference exchange and correlation quantities evaluated in
    the cc-pVTZ basis.
    (a) Magnitude of the exact-exchange energy,
    $|E_{\mathrm{x}}^{\mathrm{cc\text{-}pVTZ}}|$, as a function of the
    number of occupied orbitals,
    $N_{\mathrm{occ}}=N_e/2$, for the closed-shell species in XC100.
    (b) Distribution of the kinetic-correlation fraction,
    $T_{\mathrm{c}}^{\mathrm{cc\text{-}pVTZ}}/
    |E_{\mathrm{c}}^{\mathrm{cc\text{-}pVTZ}}|$.
    The dashed vertical line marks the median value of 0.716.}
    \label{fig:exchange_tc_fraction}
\end{figure}

Figure~\ref{fig:exchange_tc_fraction} provides additional analysis of the cc-pVTZ exchange and correlation quantities reported in XC100. 

The exact-exchange energies increase broadly in magnitude with $N_{\mathrm{occ}}$, but species with the same number of occupied orbitals can differ appreciably in $E_{\mathrm{x}}$. The distribution of $T_{\mathrm{c}}/|E_{\mathrm{c}}|$, which provides a measure of the relative contribution of kinetic correlation to the total correlation energy, is concentrated near 0.7, with a median value of 0.716. Thus, for a typical species in XC100, the kinetic-correlation energy constitutes a substantial fraction of the magnitude of the total correlation energy. 

\section{Additional DFA Error Analysis: Exchange- and Correlation-Energy Errors}

\begin{figure}[H]
    \centering
    \includegraphics[width=\textwidth]{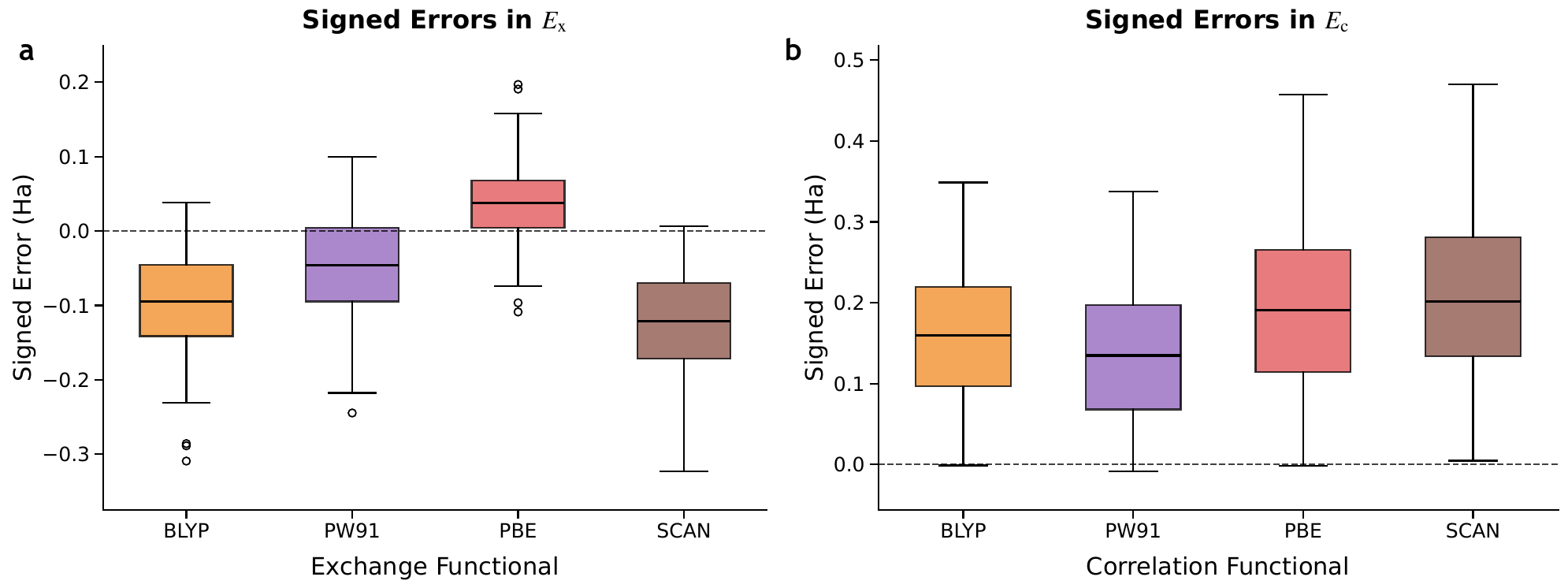}
    \caption{Distribution of signed errors in the separate exchange and
    correlation components for the selected density-functional
    approximations.
    (a) Signed errors in the exchange energy, $E_{\mathrm{x}}$, relative
    to the cc-pVTZ exact-exchange reference.
    (b) Signed errors in the correlation energy relative to the final
    composite XC100 correlation reference,
    $E_{\mathrm{c}}^{\mathrm{comp}}
    =
    E_{\mathrm{xc}}^{\mathrm{comp}}
    -
    E_{\mathrm{x}}^{\mathrm{cc\text{-}pVTZ}}$.
    Positive values indicate that the DFA value is less negative than
    the corresponding reference value, whereas negative values indicate
    that it is more negative. The horizontal dashed lines denote zero
    error. }
    \label{fig:ex_ec_signed_errors}
\end{figure}

Figure~\ref{fig:ex_ec_signed_errors} reports the signed errors in the separate exchange and correlation components for BLYP\cite{B88,LYP_1,LYP_2}, PW91\cite{PW91_1,PW91_2}, PBE\cite{PBE_1,PBE_2}, and SCAN\cite{SCAN}. Exchange errors are evaluated relative to the cc-pVTZ exact-exchange reference, whereas correlation errors are evaluated relative to the final composite XC100 correlation energy,
\[
E_{\mathrm{c}}^{\mathrm{comp}}
=
E_{\mathrm{xc}}^{\mathrm{comp}}
-
E_{\mathrm{x}}^{\mathrm{cc\text{-}pVTZ}}.
\]
Thus, the correlation-energy comparison uses the composite reference including the cc-pVQZ, Riemann CBS, and core-valence cc-pCVTZ corrections, rather than the cc-pVTZ correlation energy alone.

The exchange and correlation components exhibit distinct error patterns. BLYP, PW91 and SCAN exchange are predominantly more negative than the cc-pVTZ exact-exchange reference, while PBE exchange is shifted toward positive signed errors. In contrast, the correlation-energy errors are predominantly positive for all four functionals relative to the composite correlation reference, indicating correlation energies that are generally less negative than $E_{\mathrm{c}}^{\mathrm{comp}}$. PW91 exhibits the smallest median correlation error among the functionals shown, whereas PBE and SCAN have larger positive median errors. The differing signs of the exchange and correlation errors also demonstrate that error cancellation between the two components can contribute to lower total $E_{\mathrm{xc}}$ errors.

\section{Machine-Learned Functional Calculations and SCF Convergence}

NNGGA\cite{kanungo2025} and Skala\cite{Luise2026} were evaluated self-consistently in the cc-pVQZ basis using the same PySCF\cite{PySCF} level-9 grid as conventional DFAs. Both machine-learned functionals were first run with an SCF energy convergence threshold of $1\times10^{-10}$ Ha and a maximum of 500 iterations.


NNGGA was evaluated using the implementation of Kanungo \textit{et al.}\cite{kanungo2025} For calculations that did not converge within 500 iterations at $1\times10^{-10}$ Ha, a fresh SCF calculation was performed using a $1\times10^{-7}$ convergence threshold, again with a maximum of 500 iterations. XC100-020 (\ce{CH3F}) and XC100-092 (nitrosonium) converged using this fallback threshold and are included in the reported NNGGA statistics. Five systems remained unconverged after both attempts: XC100-003 (allene), XC100-024 (\ce{CO2}), XC100-059 (propyne), XC100-061 (acetonitrile), and XC100-093 (pyridine). Accordingly, the NNGGA $E_{\mathrm{xc}}$ analysis contains 95 XC100 species.

Skala calculations used the Skala-1.1 model with the optional D3 dispersion correction disabled.\cite{Luise2026} The primary $1\times10^{-10}$ Ha threshold and 500-iteration limit were retained without a relaxed-convergence fallback. Three systems did not converge within this protocol: XC100-067 (beryllium oxide), XC100-084 (lithium hydride), and XC100-086 (lithium oxide). Accordingly, the Skala $E_{\mathrm{xc}}$ analysis contains 97 XC100 species. Unconverged calculations were excluded from the corresponding error distributions rather than using energies from unconverged SCF computations.

\begin{table}[H]
\centering
\caption{Self-consistent convergence of the machine-learned
functionals across XC100.}
\begin{tabular}{lcl}
\hline
Functional & Converged & Unconverged XC100 systems \\
\hline
NNGGA & 95/100 &
003, 024, 059, 061, 093 \\
Skala & 97/100 &
067, 084, 086 \\
\hline
\end{tabular}
\label{tab:ml_scf_convergence}
\end{table}


\stopcontents[si]

\end{document}